\documentclass[12pt]{iopart}

\expandafter\let\csname equation*\endcsname\undefined
\expandafter\let\csname endequation*\endcsname\undefined

\usepackage{amsmath}
\usepackage{amssymb}
\usepackage{siunitx}
\usepackage{graphicx}
\usepackage{geometry}
\usepackage{xcolor}

\usepackage[%
    style=phys,%
    articletitle=true,biblabel=brackets,%
    chaptertitle=false,pageranges=false%
]{biblatex}
\DeclareFieldFormat
  [article,inbook,incollection,inproceedings,patent,thesis,unpublished]
  {title}{\mkbibemph{#1\isdot}}
\begin{document}

\title{Thermally driven elastic membranes are quasi-linear across all scales}

\author{Chanania Steinbock, Eytan Katzav}
\address{Racah Institute of Physics, The Hebrew University, Jerusalem 9190401, Israel}
\ead{chanania.steinbock@mail.huji.ac.il}

\begin{abstract}
We study the static and dynamic structure of thermally fluctuating elastic thin sheets by investigating a model known as the overdamped dynamic F\"oppl-von K\'arm\'an equation, in which the F\"oppl-von K\'arm\'an equation from elasticity theory is driven by white noise. The resulting nonlinear equation is governed by a single nondimensional coupling parameter $g$ where large and small values of $g$ correspond to weak and strong nonlinear coupling respectively. By analysing the weak coupling case with ordinary perturbation theory and the strong coupling case with a self-consistent methodology known as the self-consistent expansion, precise analytic predictions for the static and dynamic structure factors are obtained. Importantly, the maximum frequency $n_{\max}$ supported by the system plays a role in determining which of three possible classes such sheets belong to: (1) when $g \gg 1$, the system is mostly linear with roughness exponent $\zeta=1$ and dynamic exponent $z=4$, (2) when $g \ll 2/n_{\max}$, the system is extremely nonlinear with roughness exponent $\zeta=1/2$ and dynamic exponent $z=3$, (3) between these regimes, an intermediate behaviour is obtained in which a crossover occurs such that the nonlinear behaviour is observed for small frequencies while the linear behaviour is observed for large frequencies, and thus the large frequency linear tail is found to have a significant impact on the small frequency behaviour of the sheet. Back-of-the-envelope calculations suggest that ultra-thin materials such as graphene lie in this intermediate regime. Despite the existence of these three distinct behaviours, the decay rate of the dynamic structure factor is related to the static structure factor as if the system were completely linear. This quasi-linearity occurs regardless of the size of $g$ and at all length scales. Numerical simulations confirm the existence of the three classes of behaviour and the quasi-linearity of all classes.
\end{abstract}


\submitto{\jpa}
\maketitle

\section{Introduction\label{sec:intro}}

Thin matter appears at all scales in nature, ranging from everyday sheets of paper to microscopic cell membranes to geological tectonic plates. Accordingly, the theory of how thin sheets behave in the presence of thermal or uncorrelated noise has extremely wide applicability and indeed, the relatively recent synthesis of ultra-thin materials such as graphene \cite{Novoselov2005} has made the theory more relevant than ever.

The early development of such a theory \cite{NelsonBook2004} began by considering so-called ``tethered surfaces'' which were modelled by continuous elastic surfaces \cite{Kantor1986, Kantor1987}. It was found that though such sheets enter a crumpled phase at large temperatures, for sufficiently low temperatures, the thermal fluctuations actually stabilise an undulating flat phase \cite{Kantor1987a, Paczuski1988}. Using renormalisation techniques and primarily focused on the static features of such surfaces, it was found \cite{Nelson1987,  Aronovitz1989, Duplantier1990, LeDoussal1992} that the thermal fluctuations effectively renormalise the sheet's bending rigidity with the effect that its static structure factor becomes a length scale-dependent universal power-law in Fourier space. Similar results were subsequently obtained for related systems such as surfaces with intrinsic curvature, alternate geometries, inhomogeneities or exposure to tension \cite{Kosmrlj2013, Kosmrlj2014, Kosmrlj2016, Kosmrlj2017, LeDoussal2018, Shankar2021, Morshedifard2021} and were also obtained with other methods such as perturbative expansions in the membrane dimension \cite{Kardar1987}, variational perturbation theory \cite{Ahmadpoor2017} or phase field models \cite{Elder2021, Granato2022}. Simultaneously, these theoretical predictions were investigated in experimental settings \cite{Meyer2007, Meyer2007a}, including by examination of randomly crumpled surfaces \cite{PlouraboueRoux1996, BlairKudrolli2005, Balankin2006, Balankin2008, Balankin2013}, and using a variety of numerical techniques such as finite-difference methods \cite{Bilbao2008, Kosmrlj2013, Bilbao2015}, Monte Carlo simulations \cite{Bowick1996, Fasolino2007, Los2009, Troster2013a, Troster2013b} and molecular dynamics simulations \cite{Thompson2009, Thomas2015, Los2016, Ahmadpoor2017}.

While the static properties of thermally fluctuating thin sheets have been studied extensively, far less attention has been devoted to their dynamic character. Early studies of the dynamic properties of tethered membranes neglected their elastic nature \cite{Kantor1986, Kantor1987} or focused on surfaces coupled to randomly perturbing fluids \cite{Frey1991}. In recent years, detailed numerical studies \cite{Mizuochi2014, Granato2022} focused on the dynamics have begun to be performed but comprehensive theoretical descriptions remain lacking.

In this paper, we study both the static and dynamic structure of thermally fluctuating thin sheets and find them to be intimately related. Recently, we developed a model for \textit{athermally} fluctuating thin sheets with the goal of understanding the behaviour of fluctuating sheets immediately prior to deliberate crumpling \cite{Steinbock2022, Steinbock2023}. Adapting this model to provide an out-of-equilibrium model for a \textit{thermally} fluctuating thin sheet, we use a self-consistent methodology (the self-consistent expansion or SCE) to determine the sheet's static and dynamic structure factors and find among other things a previously underappreciated ``quasi-linearity'' inherent to the system. In particular, for the out-of-plane displacement field $\xi(\vec{r},t)$, the sheet's roughness exponent $\zeta$ and dynamic exponent $z$, can be defined as the characteristic exponents of the evolution of the variance of the displacement field
\begin{equation}
\left\langle \frac{1}{L^{2}}\int_{[L\times L]}d\vec{r}\,\left[\xi\left(\vec{r},t\right)-\xi\left(\vec{r},0\right)\right]^{2}\right\rangle =L^{2\zeta}f\left(\frac{t}{L^{z}}\right) \,, \label{eq:exponent definitions}
\end{equation}
where $L$ denotes the linear size of the sheet, $f(t/L^z)$ denotes some scaling function and the averaging $\langle\cdot\rangle$ denotes noise averaging. Such dynamic scaling was extensively studied in the context of growing surfaces and diffusion-limited aggregation where it is known as Family-Vicsek scaling \cite{Hohenberg1977, Vicsek1984, Family1985, BarabasiStanleyBook}.  We explicitly show that even in the presence of strong nonlinear coupling, the apparent roughness and dynamic exponents are related by the relationship
\begin{equation}
z = 2\zeta + 2 \,,
\end{equation}
which is also the expected relationship from a purely linear theory \cite{Katzav2011b, Katzav2011a}. Though such a result was previously argued for using scaling arguments \cite{Mizuochi2014}, here, we explicitly show that this quasi-linearity extends well beyond the large-scale long-wavelength behaviour and is in fact maintained at all scales.

The paper is structured as follows. In section \ref{sec:model derivation}, we outline the Langevin equation used to describe a thermally fluctuating thin sheet and frame it in terms of a Fokker-Planck equation. In section \ref{sec:linear theory}, we analyse this model for the case of weak nonlinear coupling using ordinary perturbation theory and thereby obtain the static and dynamic structure factors in sections \ref{sec:linear static structure factor} and \ref{sec:linear dynamic structure factor} respectively. In section \ref{sec:nonlinear theory}, we turn to the case of strong nonlinear coupling and, using the self-consistent expansion, derive static and dynamic structure factors in sections \ref{sec:nonlinear static structure factor} and \ref{sec:nonlinear dynamic structure factor} respectively. These analytic results are compared to numerical simulations in section 
\ref{sec:simulations} and the results are discussed in section \ref{sec:discussion}.

\section{Langevin and Fokker-Planck Equations for Fluctuating Sheets\label{sec:model derivation}}

We model fluctuating elastic thin sheets using the dynamic F\"oppl-von K\'arm\'an equations, originally investigated in the context of wave turbulence \cite{During2006, Boudaoud2008, Mordant2008, Cadot2008, Cobelli2009, Humbert2013, Miquel2013, During2015, During2017, During2019, Hassaini2019, Pavez2023} and recently further developed in \cite{Steinbock2022, Steinbock2023} to understand the structure of thin sheets driven by fluctuations immediately prior to crumpling. The basic idea is to consider a mostly flat sheet and simply write a Newton force equation for each mass element of the sheet, that is, if $\xi\left(\vec{r},t\right)$ denotes the vertical displacement of the sheet at some point relative to a reference plane with coordinates $\vec{r}=(x,y)$, then Newton's second law can generically be written as
\begin{equation}
h\rho\frac{\partial^{2}\xi}{\partial t^{2}}=-P_\mathrm{elastic}-P_\mathrm{damping}+P_\mathrm{driving}\,,\label{eq:newton equ}
\end{equation}
where $h$ and $\rho$ denote the thickness and density of the sheet respectively. The forces $P_\mathrm{driving}$ and $P_\mathrm{damping}$ correspond to external pressures which inject energy into the system and dissipate it respectively while $P_\mathrm{elastic}$ accounts for the restorative elastic character of the sheet which favours a flat undeformed configuration. Following \cite{During2006, Boudaoud2008, Mordant2008, Cadot2008, Cobelli2009, Humbert2013, Miquel2013, During2015, During2017, During2019, Hassaini2019, Steinbock2022, Steinbock2023, Pavez2023}, this restorative elastic pressure is taken to be determined by the F\"oppl-von K\'arm\'an equations \cite{LandauLifshitzElasticityBook}
\begin{align}
P_{\mathrm{elastic}} &= \kappa\nabla^{4}\xi-h\mathcal{L}(\xi,\chi) \,, \label{eq:fvk xi}\\
0 &= \nabla^{4}\chi+\frac{E}{2}\mathcal{L}(\xi,\xi) \, ,\label{eq:fvk airy}
\end{align}
where 
\begin{equation}
\kappa=\frac{Eh^{3}}{12\left(1-\nu^{2}\right)} \label{eq:bending rigidity}
\end{equation} 
denotes the bending or flexural rigidity of the sheet, $E$ and $\nu$ denote the sheet's Young's modulus and Poisson ratio respectively, $\chi\left(\vec{r},t\right)$ denotes the Airy stress potential of the sheet and $\mathcal{L}(f,g)$ denotes a bilinear symmetric operator which is given in Cartesian coordinates by
\begin{equation}
\mathcal{L}\left(f,g\right)=\frac{\partial^{2}f}{\partial x^{2}}\frac{\partial^{2}g}{\partial y^{2}}+\frac{\partial^{2}f}{\partial y^{2}}\frac{\partial^{2}g}{\partial x^{2}}-2\frac{\partial^{2}f}{\partial x\partial y}\frac{\partial^{2}g}{\partial x\partial y}\,.
\end{equation}

The F\"oppl-von K\'arm\'an elastic term describes the reduction of a three-dimensional linear-elastic thin body to a two-dimensional elastic sheet. As a three-dimensional body, the sheet is assumed to be purely elastic. In the two-dimensional reduction, the nonlinear terms introduced by the bilinear operator $\mathcal{L}$ correspond to a geometric nonlinearity in which the stretching of the material modifies the local Gaussian curvature of the sheet. This in turn is known to have the effect of focusing stresses in the sheet thus giving rise to the appearance of singular structures in crumpled sheets such as $d$-cones and ridges \cite{Lobkovsky1995, Lobkovsky1997, BenAmar1997, Witten2007}.

Various driving and damping pressures can and have been considered previously. In the context of wave turbulence, little consideration is given to the particular sources and sinks of energy beyond energy injection occurring at large scales and dissipation occurring at small scales as this is sufficient to observe an energy cascade reminiscent of hydrodynamic turbulence. A more concrete approach is considered in \cite{Steinbock2022, Steinbock2023} in which the driving pressure is taken to be conserved white noise which conserves linear momentum, while the damping pressure is taken to be ordinary fluid friction with coefficient of friction $\alpha$
\begin{equation}
P_\textrm{damping}=\alpha\frac{\partial\xi}{\partial t}\,.
\end{equation}

In this work, we adapt this latter approach by considering the same type of damping while the system is driven out-of-equilibrium by thermal noise at temperature $T$, that is, we consider driving pressures $P_\textrm{driving} = \eta\left(\vec{r},t\right)$ where 
\begin{align}
\left\langle \eta\left(\vec{r},t\right)\right\rangle &=0\,,\\
\left\langle \eta\left(\vec{r},t\right)\eta\left(\vec{r}\,',t'\right)\right\rangle &=
2\alpha k_\mathrm{B} T\delta\left(t-t'\right)\delta\left(\vec{r}-\vec{r}\,'\right)\,,
\end{align}
and $k_\mathrm{B}$ is the Boltzmann constant. Unlike the conserved noise considered in \cite{Steinbock2022, Steinbock2023} which is an appropriate model when the noise originates from forces internal to the sheet, as would occur for example with active sheets \cite{Cagnetta2022}, the thermal noise considered here is an appropriate model for a sheet driven by external forces such as coupling to a thermal bath, for example, a surrounding fluid.

For a sheet with dimensions $L\times L$, the vertical displacement $\xi(\vec{r},t)$, Airy stress potential $\chi(\vec{r},t)$ and noise $\eta(\vec{r},t)$ can be written in terms of their of Fourier components $\tilde{\xi}_{\vec{n}}(t)$, $\tilde{\chi}_{\vec{n}}(t)$ and $\tilde{\eta}_{\vec{n}}(t)$ as
\begin{align}
\xi(\vec{r},t) &= \sum_{\vec{n}\in\mathbb{Z}^2}\tilde{\xi}_{\vec{n}}(t)e^{i\frac{2\pi}{L}\vec{n}\cdot\vec{r}}\, , \\
\tilde{\xi}_{\vec{n}}(t) &= \frac{1}{L^2}\int_{[L\times L]}d\vec{r}\,\xi(\vec{r},t)e^{-i\frac{2\pi}{L}\vec{n}\cdot\vec{r}}
\end{align}
and similarly for $\chi(\vec{r},t)$ and $\eta(\vec{r},t)$. Writing the F\"oppl-von K\'arm\'an equations in terms of these Fourier components and substituting the above relationships into equation~(\ref{eq:newton equ}) allows us to write a single nonlinear Langevin equation for the Fourier components $\tilde{\xi}_{\vec{n}}(t)$ of the vertical displacement. For the sake of convenience, we will only consider the overdamped limit here such that the left-hand side of equation (\ref{eq:newton equ}) can be neglected and thus we obtain the nonlinear Langevin equation
\begin{equation}
\alpha\frac{\partial\tilde{\xi}_{\vec{n}}}{\partial t}=-\frac{\left(2\pi\right)^{4}\kappa}{L^4}\left|\vec{n}\right|^{4}\tilde{\xi}_{\vec{n}}-\frac{\left(2\pi\right)^{4}Eh}{2L^4}\sum_{\vec{\ell}_{1}\ne\vec{n}}\sum_{\vec{\ell}_{2}}\sum_{\vec{\ell}_{3}}V_{\vec{n},\vec{\ell}_{1},\vec{\ell}_{2},\vec{\ell}_{3}}\tilde{\xi}_{\vec{\ell}_{1}}\tilde{\xi}_{\vec{\ell}_{2}}\tilde{\xi}_{\vec{\ell}_{3}}+\tilde{\eta}_{\vec{n}} \label{eq:langevin equ dimen}
\end{equation}
where the kernel of the sum, $V_{\vec{n},\vec{\ell}_{1},\vec{\ell}_{2},\vec{\ell}_{3}}$, is given by
\begin{equation}
V_{\vec{n},\vec{\ell}_{1},\vec{\ell}_{2},\vec{\ell}_{3}}=\delta_{\vec{n},\vec{\ell}_{1}+\vec{\ell}_{2}+\vec{\ell}_{3}}\frac{\left|\vec{n}\times\vec{\ell}_{1}\right|^{2}\left|\vec{\ell}_{2}\times\vec{\ell}_{3}\right|^{2}}{\left|\vec{n}-\vec{\ell}_{1}\right|^{4}} \label{eq:V}
\end{equation}
and each sum is taken over all lattice points of $\mathbb{Z}^2$ excluding the point $\vec{\ell}_1=\vec{n}$. The variance of the noise in Fourier space is given by
\begin{equation}
\left\langle \tilde{\eta}_{\vec{n}}\left(t\right)\tilde{\eta}_{\vec{n}'}\left(t'\right)\right\rangle =\frac{2\alpha k_{\mathrm{B}}T}{L^{2}}\delta\left(t-t'\right)\delta_{\vec{n},-\vec{n}'} \,.
\end{equation}

Formally, equation~(\ref{eq:langevin equ dimen}) is a nonlinear Langevin equation with a quartic interaction. Accordingly, it can be thought of as a type of $\phi^{4}$-field theory \cite{Kleinert2001} with a spatially varying kernel $V_{\vec{n},\vec{\ell}_{1},\vec{\ell}_{2},\vec{\ell}_{3}}$. Importantly, the behaviour of the zeroth mode $\tilde{\xi}_{0}$ is completely decoupled from that of the other modes since subbing $\vec{n}=0$ into equation~(\ref{eq:langevin equ dimen}) gives
\begin{equation}
\alpha\frac{\partial\tilde{\xi}_{0}}{\partial t}=\tilde{\eta}_{0}(t) \label{eq: zeroth mode equ}
\end{equation}
while the kernel $V_{\vec{n},\vec{\ell}_{1},\vec{\ell}_{2},\vec{\ell}_{3}}$ vanishes whenever $\vec{\ell}_1$, $\vec{\ell}_2$ or $\vec{\ell}_3$ equal 0 thus ensuring that for all $\vec{n}\ne0$, the mode $\tilde{\xi}_{0}$ provides no contribution to the Langevin equation. This has the strong implication that the mean position of the sheet, which is given by the zeroth mode $\tilde{\xi}_{0}$, simply performs a random walk in space with no effect on the structure of the sheet and the sums over $\vec{\ell}_1$, $\vec{\ell}_2$ or $\vec{\ell}_3$ can be taken as skipping over the origin.

The structure of a sheet undergoing thermal fluctuations described by equation~(\ref{eq:langevin equ dimen}) is most easily approached by studying its corresponding Fokker-Planck equation for the probability $P=P(\{ \tilde{\xi}_{\vec{n}}(t)\},t)$ of finding the sheet with Fourier components $\{ \tilde{\xi}_{\vec{n}}(t)\}$ at time $t$ \cite{RiskenBook}
\begin{multline}
\frac{\partial P}{\partial t}=\frac{k_{\mathrm{B}}T}{\alpha L^{2}}\sum_{\vec{n}}\frac{\partial^{2}P}{\partial\tilde{\xi}_{\vec{n}}\partial\tilde{\xi}_{-\vec{n}}} +\frac{\left(2\pi\right)^{4}\kappa}{\alpha L^{4}}\sum_{\vec{n}}\left|\vec{n}\right|^{4}\frac{\partial}{\partial\tilde{\xi}_{\vec{n}}}\left[\tilde{\xi}_{\vec{n}}P\right] \\
+\frac{\left(2\pi\right)^{4}Eh}{2\alpha L^{4}}\sum_{\vec{n}}\frac{\partial}{\partial\tilde{\xi}_{\vec{n}}}\Biggl[P\sum_{\begin{subarray}{c}
\vec{\ell}_{1}\ne\vec{n}\\
\vec{\ell}_{2},\vec{\ell}_{3}
\end{subarray}}V_{\vec{n},\vec{\ell}_{1},\vec{\ell}_{2},\vec{\ell}_{3}}\tilde{\xi}_{\vec{\ell}_{1}}\tilde{\xi}_{\vec{\ell}_{2}}\tilde{\xi}_{\vec{\ell}_{3}}\Biggr]\,. \label{eq:FP equ dimen}
\end{multline}
Multiplying this equation by any function $\mathbb{F}(\{\tilde{\xi}_{\vec{n}}(t)\})$ of the Fourier components and integrating over all components $\tilde{\xi}_{\vec{n}}(t)$ gives after some integration by parts
\begin{multline}
\frac{\partial\left\langle \mathbb{F}\right\rangle }{\partial t}=\frac{k_{\mathrm{B}}T}{\alpha L^{2}}\sum_{\vec{n}}\left\langle \frac{\partial^{2}\mathbb{F}}{\partial\tilde{\xi}_{\vec{n}}\partial\tilde{\xi}_{-\vec{n}}}\right\rangle -\frac{\left(2\pi\right)^{4}\kappa}{\alpha L^{4}}\sum_{\vec{n}}\left|\vec{n}\right|^{4}\left\langle \frac{\partial\mathbb{F}}{\partial\tilde{\xi}_{\vec{n}}}\tilde{\xi}_{\vec{n}}\right\rangle \\
-\frac{\left(2\pi\right)^{4}Eh}{2\alpha L^{4}}\sum_{\vec{n}}\sum_{\begin{subarray}{c}
\vec{\ell}_{1}\ne\vec{n}\\
\vec{\ell}_{2},\vec{\ell}_{3}
\end{subarray}}V_{\vec{n},\vec{\ell}_{1},\vec{\ell}_{2},\vec{\ell}_{3}}\left\langle \frac{\partial\mathbb{F}}{\partial\tilde{\xi}_{\vec{n}}}\tilde{\xi}_{\vec{\ell}_{1}}\tilde{\xi}_{\vec{\ell}_{2}}\tilde{\xi}_{\vec{\ell}_{3}}\right\rangle \, , \label{eq:moment equ dimen}
\end{multline}
where we have defined the averages
\begin{equation}
\left\langle \mathbb{F}\right\rangle =\int\prod_{\vec{n}}d\tilde{\xi}_{\vec{n}}\,\mathbb{F}(\{\tilde{\xi}_{\vec{n}}(t)\})P(\{\tilde{\xi}_{\vec{n}}(t)\},t) \, .
\end{equation}

The first term on the right-hand side of equations~(\ref{eq:FP equ dimen}) and (\ref{eq:moment equ dimen}) corresponds to diffusion through the phase space of the fluctuating sheet due to the thermal noise while the second term corresponds to the limitation of this diffusion due to the restorative effect of the sheet's linear elasticity. The final term of these equations modifies this restorative effect by accounting for the geometric nonlinearity of a deformed sheet as described by the nonlinear portion of the F\"oppl-von K\'arm\'an equations (equations~(\ref{eq:fvk xi}) and (\ref{eq:fvk airy})) and its corresponding kernel in Fourier space $V_{\vec{n},\vec{\ell}_{1},\vec{\ell}_{2},\vec{\ell}_{3}}$. In instances where the effect of the geometric nonlinearity is weak, the final term can be neglected at zeroth order and a linear theory is obtained which in principle can be perturbed around. In contrast, when the effect of the nonlinearity is large, the linear theory fails to provide any sort of reasonable approximation of the sheet and more sophisticated techniques of analysis must be employed. In the next sections, we will argue that both of these regimes are relevant, interesting and worth addressing. We will begin by analysing the behaviour of the sheet in its linear regime before moving onto its nonlinear regime.

\section{The Linear Regime\label{sec:linear theory}}

To identify the linear regime, it is helpful to begin by nondimensionalising our equations. Examining the physical parameters which enter our model, we find that one can define the dimensionless quantity
\begin{equation}
g=\frac{\left(2\pi\right)^{2}\kappa}{\sqrt{2k_{\mathrm{B}}TEhL^{2}}} \label{eq:g definition}
\end{equation}
where the size of $g$ will determine whether we are in the linear or nonlinear regime. If we define dimensionless scalings of the time $\hat{t}$ and vertical displacement $\hat{\xi}$ as follows,
\begin{align}
\hat{t}&=\frac{\left(2\pi\right)^{4}\kappa}{\alpha L^{4}}t \,,\\
\hat{\xi}_{\vec{n}}\left(\hat{t}\,\right) &=\left[\frac{\left(2\pi\right)^{4}\kappa}{2k_{\mathrm{B}}TL^{2}}\right]^{1/2}\tilde{\xi}_{\vec{n}}\left(t\right) \,,
\end{align}
then equation~(\ref{eq:moment equ dimen}) can be written as
\begin{multline}
\frac{\partial\left\langle \mathbb{F}\right\rangle }{\partial\hat{t}}=\frac{1}{2}\sum_{\vec{n}}\left\langle \frac{\partial^{2}\mathbb{F}}{\partial\hat{\xi}_{\vec{n}}\partial\hat{\xi}_{-\vec{n}}}\right\rangle -\sum_{\vec{n}}\left|\vec{n}\right|^{4}\left\langle \frac{\partial\mathbb{F}}{\partial\hat{\xi}_{\vec{n}}}\hat{\xi}_{\vec{n}}\right\rangle \\
-\frac{1}{2}\frac{1}{g^{2}}\sum_{\vec{n}}\sum_{\begin{subarray}{c}
\vec{\ell}_{1}\ne\vec{n}\\
\vec{\ell}_{2},\vec{\ell}_{3}
\end{subarray}}V_{\vec{n},\vec{\ell}_{1},\vec{\ell}_{2},\vec{\ell}_{3}}\left\langle \frac{\partial\mathbb{F}}{\partial\hat{\xi}_{\vec{n}}}\hat{\xi}_{\vec{\ell}_{1}}\hat{\xi}_{\vec{\ell}_{2}}\hat{\xi}_{\vec{\ell}_{3}}\right\rangle \,, \label{eq:moment eq nondim}
\end{multline}
which contains just the single dimensionless parameter $g$ and thus we find that we will be in the linear regime if $g \gg 1$.

To understand the physical significance of the dimensionless parameter $g$, it is helpful to recast equation~(\ref{eq:g definition}) in terms which make the dependence on the physical size explicit. In particular, it is known from equation~(\ref{eq:bending rigidity}) that the flexural rigidity $\kappa$ grows with the sheet thickness as $\sim h^3$. Accordingly, in terms of the Young's modulus $E$ and Poisson ratio $\nu$, which are both independent of sheet thickness, $g$ can be written as
\begin{equation}
g=\frac{\left(2\pi\right)^{2}}{12(1-\nu^2)}\sqrt{\frac{E h^5}{2k_{\mathrm{B}}TL^{2}}} \,. \label{eq:g definition 2}
\end{equation}
This expression makes it clear that $g$ grows with sheet thickness and thus nonlinear effects will become more important for thinner sheets.

Physically, the condition $g \gg 1$ occurs for macroscopic sheets at ordinary temperatures. For instance, a sheet of aluminium with length $L\sim20\textrm{ cm}$ to a side, would have a typical thickness around $h\sim0.02\textrm{ mm}$, a Young's modulus around $E\sim70\textrm{ GPa}$, a Poisson ratio of $\nu\sim0.35$ and thus a bending rigidity $\kappa\sim5\times10^{-5}\textrm{ N m}$. Accordingly, at temperatures of say $T\sim300\textrm{ K}$, we would have $g\sim10^5$ which is indeed much much larger than unity. 

For such sheets, the F\"oppl-von K\'arm\'an geometric nonlinearity can be completely neglected and if we so desire, a perturbative expansion can be carried out around the linear part. To this end, let $\left\langle \mathbb{F}\right\rangle^{(m)}$ denote an $m^\textrm{th}$ order approximation for any quantity $\left\langle \mathbb{F}\right\rangle$, that is, $\left\langle \mathbb{F}\right\rangle^{(m)}$ denotes an expansion containing all corrections up to order $m$. Then the following iterative scheme
\begin{multline}
\frac{\partial\left\langle \mathbb{F}\right\rangle ^{\left(m\right)}}{\partial\hat{t}}=\frac{1}{2}\sum_{\vec{n}}\left\langle \frac{\partial^{2}\mathbb{F}}{\partial\hat{\xi}_{\vec{n}}\partial\hat{\xi}_{-\vec{n}}}\right\rangle ^{\left(m\right)}-\sum_{\vec{n}}\left|\vec{n}\right|^{4}\left\langle \frac{\partial\mathbb{F}}{\partial\hat{\xi}_{\vec{n}}}\hat{\xi}_{\vec{n}}\right\rangle ^{\left(m\right)}\\-\frac{1}{2}\frac{1}{g^{2}}\sum_{\vec{n}}\sum_{\begin{subarray}{c}
\vec{\ell}_{1}\ne\vec{n}\\
\vec{\ell}_{2},\vec{\ell}_{3}
\end{subarray}}V_{\vec{n},\vec{\ell}_{1},\vec{\ell}_{2},\vec{\ell}_{3}}\left\langle \frac{\partial\mathbb{F}}{\partial\hat{\xi}_{\vec{n}}}\hat{\xi}_{\vec{\ell}_{1}}\hat{\xi}_{\vec{\ell}_{2}}\hat{\xi}_{\vec{\ell}_{3}}\right\rangle ^{\left(m-1\right)} \label{eq:moment eq nondim linear}
\end{multline}
can be used to calculate any expectation value $\left\langle \mathbb{F}\right\rangle$ order by order. In this notation, $\left\langle \mathbb{F}\right\rangle^{(0)}$ corresponds to the solution to the purely linear theory and since the small parameter of equation~(\ref{eq:moment eq nondim linear}) is $1/g^2$, $\left\langle \mathbb{F}\right\rangle^{(m)}$ will contain a term proportional to $1/g^{2m}$. In the following subsections, we demonstrate this approach by calculating the static and dynamic structure factors up to first order in the linear regime.

\subsection{The Static Structure Factor\label{sec:linear static structure factor}}

The static structure factor $\langle \hat{\xi}_{\vec{n}}\hat{\xi}_{-\vec{n}} \rangle $ can be calculated at zeroth order by subbing $\mathbb{F}=\hat{\xi}_{\vec{n}_{1}}\hat{\xi}_{\vec{n}_{2}}$ and $m=0$ into equation~(\ref{eq:moment eq nondim linear})
\begin{equation}
0=\frac{1}{2}\sum_{\vec{n}}\left\langle \frac{\partial^{2}\left(\hat{\xi}_{\vec{n}_{1}}\hat{\xi}_{\vec{n}_{2}}\right)}{\partial\hat{\xi}_{\vec{n}}\partial\hat{\xi}_{-\vec{n}}}\right\rangle ^{\left(0\right)} -\sum_{\vec{n}}\left|\vec{n}\right|^{4}\left\langle \frac{\partial\left(\hat{\xi}_{\vec{n}_{1}}\hat{\xi}_{\vec{n}_{2}}\right)}{\partial\hat{\xi}_{\vec{n}}}\hat{\xi}_{\vec{n}}\right\rangle ^{\left(0\right)} \,.
\end{equation}
Note that the left-hand side has been set to 0 since in steady-state, the equal-time two-point function $\langle \hat{\xi}_{\vec{n}_{1}}\hat{\xi}_{\vec{n}_{2}}\rangle$ does not depend on time. Calculating the derivatives and simplifying the sums, we obtain the static two-point function
\begin{equation}
\left\langle \hat{\xi}_{\vec{n}_{1}}\hat{\xi}_{\vec{n}_{2}}\right\rangle ^{\left(0\right)}=\frac{\delta_{\vec{n}_{1},-\vec{n}_{2}}}{\left|\vec{n}_{1}\right|^{4}+\left|\vec{n}_{2}\right|^{4}}=\frac{\delta_{\vec{n}_{1},-\vec{n}_{2}}}{2\left|\vec{n}_{1}\right|^{4}} \, . \label{eq:linear stat 2pt zeroth order}
\end{equation}

If we wish to consider macroscopic sheets being driven by white noise of arbitrary amplitude, i.e. by effective temperatures much larger than room temperature and thus with smaller corresponding values of $g$, corrections to this expression may become relevant. To this end, subbing $\mathbb{F}=\hat{\xi}_{\vec{n}_{1}}\hat{\xi}_{\vec{n}_{2}}$ and $m=1$ into equation~(\ref{eq:moment eq nondim linear}) gives, after some simplification, the equation
\begin{multline}
0=\delta_{\vec{n}_{1},-\vec{n}_{2}}-\left(\left|\vec{n}_{1}\right|^{4}+\left|\vec{n}_{2}\right|^{4}\right)\left\langle \hat{\xi}_{\vec{n}_{1}}\hat{\xi}_{\vec{n}_{2}}\right\rangle ^{\left(1\right)}\\-\frac{1}{2}\frac{1}{g^{2}}\Biggl[\, \sum_{\begin{subarray}{c}
\vec{\ell}_{1}\ne\vec{n}_{1}\\
\vec{\ell}_{2},\vec{\ell}_{3}
\end{subarray}}V_{\vec{n}_{1},\vec{\ell}_{1},\vec{\ell}_{2},\vec{\ell}_{3}}\left\langle \hat{\xi}_{\vec{n}_{2}}\hat{\xi}_{\vec{\ell}_{1}}\hat{\xi}_{\vec{\ell}_{2}}\hat{\xi}_{\vec{\ell}_{3}}\right\rangle ^{\left(0\right)} +\sum_{\begin{subarray}{c}
\vec{\ell}_{1}\ne\vec{n}_{2}\\
\vec{\ell}_{2},\vec{\ell}_{3}
\end{subarray}}V_{\vec{n}_{2},\vec{\ell}_{1},\vec{\ell}_{2},\vec{\ell}_{3}}\left\langle \hat{\xi}_{\vec{n}_{1}}\hat{\xi}_{\vec{\ell}_{1}}\hat{\xi}_{\vec{\ell}_{2}}\hat{\xi}_{\vec{\ell}_{3}}\right\rangle ^{\left(0\right)}\Biggr]\,. \label{eq:linear 2pt func derivation}
\end{multline}
To proceed, the zeroth order equal-time four-point function $\langle \hat{\xi}_{\vec{n}_{1}}\hat{\xi}_{\vec{\ell}_{1}}\hat{\xi}_{\vec{\ell}_{2}}\hat{\xi}_{\vec{\ell}_{3}}\rangle ^{\left(0\right)}$ is required. This can be obtained by subbing $\mathbb{F}=\hat{\xi}_{\vec{n}_1}\hat{\xi}_{\vec{\ell}_{1}}\hat{\xi}_{\vec{\ell}_{2}}\hat{\xi}_{\vec{\ell}_{3}}$ and $m=0$ into equation~(\ref{eq:moment eq nondim linear}), which after some straightforward manipulation results in
\begin{multline}
\left\langle \hat{\xi}_{\vec{n}_{1}}\hat{\xi}_{\vec{\ell}_{1}}\hat{\xi}_{\vec{\ell}_{2}}\hat{\xi}_{\vec{\ell}_{3}}\right\rangle ^{\left(0\right)}=\left\langle \hat{\xi}_{\vec{n}_{1}}\hat{\xi}_{\vec{\ell}_{1}}\right\rangle ^{\left(0\right)}\left\langle \hat{\xi}_{\vec{\ell}_{2}}\hat{\xi}_{\vec{\ell}_{3}}\right\rangle ^{\left(0\right)}\\+\left\langle \hat{\xi}_{\vec{n}_{1}}\hat{\xi}_{\vec{\ell}_{2}}\right\rangle ^{\left(0\right)}\left\langle \hat{\xi}_{\vec{\ell}_{1}}\hat{\xi}_{\vec{\ell}_{3}}\right\rangle ^{\left(0\right)}+\left\langle \hat{\xi}_{\vec{n}_{1}}\hat{\xi}_{\vec{\ell}_{3}}\right\rangle ^{\left(0\right)}\left\langle \hat{\xi}_{\vec{\ell}_{1}}\hat{\xi}_{\vec{\ell}_{2}}\right\rangle ^{\left(0\right)} \label{eq:linear stat 4pt zeroth order}
\end{multline}
This is of course none other than Isserlis' theorem \cite{Isserlis1916, Isserlis1918}, also known as Wick's theorem \cite{Kardar2007Book}, for the four-point function of a multivariate Gaussian distribution. Accordingly, combining this result and the definition of $V_{\vec{n}_{2},\vec{\ell}_{1},\vec{\ell}_{2},\vec{\ell}_{3}}$ given by equation~(\ref{eq:V}) allows us to obtain
\begin{equation}
\sum_{\begin{subarray}{c}
\vec{\ell}_{1}\ne\vec{n}_{2}\\
\vec{\ell}_{2},\vec{\ell}_{3}
\end{subarray}}V_{\vec{n}_{2},\vec{\ell}_{1},\vec{\ell}_{2},\vec{\ell}_{3}}\left\langle \hat{\xi}_{\vec{n}_{1}}\hat{\xi}_{\vec{\ell}_{1}}\hat{\xi}_{\vec{\ell}_{2}}\hat{\xi}_{\vec{\ell}_{3}}\right\rangle ^{\left(0\right)}=\left\langle \hat{\xi}_{\vec{n}_{1}}\hat{\xi}_{\vec{n}_{2}}\right\rangle ^{\left(0\right)}\sum_{\vec{\ell}_{1}\ne\vec{n}_{2}}\frac{\left|\vec{n}_{2}\times\vec{\ell}_{1}\right|^{4}}{\left|\vec{\ell}_{1}\right|^{4}\left|\vec{n}_{2}-\vec{\ell}_{1}\right|^{4}} \label{eq:V sum}
\end{equation}
and thus equation~(\ref{eq:linear 2pt func derivation}) can be rearranged to give the static structure factor as
\begin{equation}
\left\langle \hat{\xi}_{\vec{n}_{1}}\hat{\xi}_{\vec{n}_{2}}\right\rangle ^{\left(1\right)}=\left\langle \hat{\xi}_{\vec{n}_{1}}\hat{\xi}_{\vec{n}_{2}}\right\rangle ^{\left(0\right)}-\frac{1}{g^{2}}\left(\left\langle \hat{\xi}_{\vec{n}_{1}}\hat{\xi}_{\vec{n}_{2}}\right\rangle ^{\left(0\right)}\right)^{2}\sum_{\vec{\ell}\ne\vec{n}_{1}}\frac{\left|\vec{n}_{1}\times\vec{\ell}\,\right|^{4}}{\left|\vec{\ell}\,\right|^{4}\left|\vec{n}_{1}-\vec{\ell}\,\right|^{4}}\,.
\end{equation}
Though the summand in this expression is ill defined when $\vec{\ell}=(0,0)$, recall from section \ref{sec:model derivation} that no harm is done when the sums skip over the origin since the zeroth mode $\tilde{\xi}_0$ is completely decoupled from the other modes anyway. The sum can be quickly evaluated by approximating it as the integral
\begin{equation}
\sum_{\vec{\ell}\ne0,\vec{n}}\frac{\left|\vec{n}\times\vec{\ell}\,\right|^{4}}{\left|\vec{\ell}\,\right|^{4}\left|\vec{n}-\vec{\ell}\,\right|^{4}}\approx\int d^{2}\ell\,\frac{\left|\vec{n}\times\vec{\ell}\,\right|^{4}}{\left|\vec{\ell}\,\right|^{4}\left|\vec{n}-\vec{\ell}\,\right|^{4}} \,. \label{eq:sum integ approx}
\end{equation}
Noting that $|\vec{n}\times\vec{\ell}\, |^{4} = | \vec{n} |^4 | \vec{\ell} \, |^4\sin^4\theta$ where $\theta$ is the angle between $\vec{n}$ and $\vec{\ell}$ allows this integral to be calculated in polar coordinates. Writing the integral as
\begin{equation}
\int d^{2}\ell\,\frac{\left|\vec{n}\times\vec{\ell}\,\right|^{4}}{\left|\vec{\ell}\,\right|^{4}\left|\vec{n}-\vec{\ell}\,\right|^{4}}=\\
n^{4}\int_{0}^{\infty}d\ell\,\ell\int_{0}^{2\pi}d\theta\frac{\sin^{4}\theta}{\left(n^{2}+\ell^{2}-2n\ell\cos\theta\right)^{2}} \label{eq:explicit integral}
\end{equation}
and noting that the angular integral contributes
\begin{equation}
\int_{0}^{2\pi}d\theta\,\frac{\sin^{4}\theta}{\left(n^{2}+\ell^{2}-2n\ell\cos\theta\right)^{2}}=\frac{3\pi}{4}\begin{cases}
n^{-4} & \ell\le n\\
\ell^{-4} & \ell\ge n
\end{cases} \,, \label{eq:angular integral}
\end{equation}
we find that
\begin{equation}
\int d^{2}\ell\,\frac{\left|\vec{n}\times\vec{\ell}\,\right|^{4}}{\left|\vec{\ell}\,\right|^{4}\left|\vec{n}-\vec{\ell}\,\right|^{4}} = \frac{3\pi}{4}n^2\,. \label{eq:evaluated integral}
\end{equation}
Accordingly, for large $g$, the static structure factor $\langle \hat{\xi}_{\vec{n}}\hat{\xi}_{-\vec{n}}\rangle$ is given up to first order by
\begin{equation}
\left\langle \hat{\xi}_{\vec{n}}\hat{\xi}_{-\vec{n}}\right\rangle ^{\left(1\right)}=\frac{1}{2n^{4}}\left(1-\frac{3\pi}{8g^{2}}\frac{1}{n^{2}}\right) \,.
\end{equation}
Equivalently, we can rewrite this as the straightforward Pad\'e approximant \cite{PadeApproximantsBook1996}
\begin{equation}
\left\langle \hat{\xi}_{\vec{n}}\hat{\xi}_{-\vec{n}}\right\rangle ^{\left(1\right)}=\frac{1}{2n^{4}\left(1+\frac{3\pi}{8g^{2}}\frac{1}{n^{2}}\right)} \,, \label{eq:pade linear}
\end{equation}
which makes it clear that the first order correction is simply a renormalisation of the linear coupling $n^4$ by $n^{4}\left(1+\frac{3\pi}{8g^{2}}\frac{1}{n^{2}}\right)$. Calculating higher order corrections is of course possible though tedious.

Returning to dimensional quantities, we find that
\begin{equation}
\left\langle \tilde{\xi}_{\vec{n}}\tilde{\xi}_{-\vec{n}}\right\rangle ^{\left(1\right)}=\frac{k_{\mathrm{B}}TL^{2}}{\left(2\pi\right)^{4}\kappa n^{4}\left(1+\frac{3\pi}{8\left(2\pi\right)^{4}}\frac{2k_{\mathrm{B}}TEhL^{2}}{\kappa^{2}}\frac{1}{n^{2}}\right)} \,. \label{eq:lin stat 2 pt dimensional}
\end{equation}

The scaling picture described by equation~(\ref{eq:exponent definitions}) can be translated to Fourier space \cite{Katzav2004} where it implies that the structure factor decays as
\begin{equation}
\left\langle \tilde{\xi}_{\vec{n}}\tilde{\xi}_{-\vec{n}}\right\rangle \sim\frac{1}{n^{d+2\zeta}} \,, \label{eq:zeta definition}
\end{equation}
where $\zeta$ denotes the roughness exponent of the sheet and $d$ its dimension. We have $d=2$ and thus comparing this definition of the roughness exponent with the result presented by equation~(\ref{eq:lin stat 2 pt dimensional}), it is clear that the roughness exponent of the linear regime is simply $\zeta=1$ though this only characterises the very large scale behaviour of the sheet.

Finally, when addressing the nonlinear regime, it will become apparent that, under certain conditions, systems which have an upper frequency cut-off can behave differently from those which don't. Accordingly, it is worth considering the effect of an upper cut-off already at this stage. Conceptually, suppose that at sufficiently fine scales, the system has some minimum length scale $a\ll L$ beneath which the F\"oppl-von K\'arm\'an equations no longer apply. This introduces a new dimensionless quantity into our system, $n_\mathrm{max} = L/(2a) \gg 1$, which enters our calculations as an upper bound on the sum and integral given in equation~(\ref{eq:sum integ approx}). Replacing the $\infty$ in the upper-bound of equation~(\ref{eq:explicit integral}) with the upper frequency cut-off $n_\mathrm{max}$, we ultimately find that
\begin{equation}
\left\langle \hat{\xi}_{\vec{n}}\hat{\xi}_{-\vec{n}}\right\rangle ^{\left(1\right)}=\frac{1}{2n^{4}\left(1+\frac{3\pi}{8g^{2}}\frac{1}{n^{2}}\left[1-\frac{1}{2}\left(\frac{n}{n_{\mathrm{max}}}\right)^{2}\right]\right)} \,. \label{eq:lin stat 2 pt w cutoff}
\end{equation}
That is, the existence of an upper frequency cut-off has the effect of slightly modifying the first correction to the renormalised linear coupling. Notice that in the limit of $n_\mathrm{max}\rightarrow\infty$, we simply return to the result described by equation~(\ref{eq:pade linear}) and thus the upper cut-off has little effect in the linear regime. Furthermore, for small $n$ (i.e. $n\sim 1$), the effect of the upper cut-off is extremely small and thus the expression given by equation~(\ref{eq:pade linear}) is perfectly sufficient, while for large $n$ (i.e. $n\sim n_\mathrm{max}$), the effect of the upper cut-off on the first order correction might indeed be large but since the entire first order correction becomes increasingly small for large $n$, the effect of the upper cut-off is again negligible. We thus find that in the linear regime, an upper frequency cut-off can be safely neglected.

\subsection{The Dynamic Structure Factor \label{sec:linear dynamic structure factor}}

Equation~(\ref{eq:moment eq nondim linear}) can also be used to calculate dynamic quantities such as the dynamic structure factor $\langle \hat{\xi}_{\vec{n}}(0)\hat{\xi}_{-\vec{n}}(\hat{t}\,)\rangle$, though first, we should clarify a notational ambiguity. The derivatives in equation~(\ref{eq:moment eq nondim linear}) are with respect to $\hat{\xi}_{\vec{n}}=\hat{\xi}_{\vec{n}}(\hat{t}\,)$, that is, for the sake of conciseness, the time dependence has been implicitly understood to exist, even if it is not noted explicitly. Until now, this made little difference as the static quantities considered were independent of time such that $\langle \hat{\xi}_{\vec{n}}(\hat{t}\,)\hat{\xi}_{-\vec{n}}(\hat{t}\,) \rangle= \langle \hat{\xi}_{\vec{n}}(0)\hat{\xi}_{-\vec{n}}(0) \rangle$ and similarly for the static four-point function. For time-dependent quantities however, the time dependence cannot be ignored thus while we will continue to use this notation, we will need to bear in mind that if unspecified, $\hat{\xi}_{\vec{n}}$ is merely shorthand for $\hat{\xi}_{\vec{n}}(\hat{t}\,)$. If some other time such as $\hat{\xi}_{\vec{n}}(0)$ is intended, this will always be stated explicitly.

Now, subbing in $\mathbb{F}=\hat{\xi}_{\vec{n}_{1}}\left(0\right)\hat{\xi}_{\vec{n}_{2}}(\hat{t}\,)$ and $m=0$ into equation~(\ref{eq:moment eq nondim linear}) results in the ODE
\begin{equation}
\frac{\partial}{\partial\hat{t}}\left\langle \hat{\xi}_{\vec{n}_{1}}\left(0\right)\hat{\xi}_{\vec{n}_{2}}(\hat{t}\,)\right\rangle ^{\left(0\right)}=-\left|\vec{n}_{2}\right|^{4}\left\langle \hat{\xi}_{\vec{n}_{1}}\left(0\right)\hat{\xi}_{\vec{n}_{2}}(\hat{t}\,)\right\rangle ^{\left(0\right)} \,,
\end{equation}
which is simply solved by
\begin{equation}
\left\langle \hat{\xi}_{\vec{n}_{1}}\left(0\right)\hat{\xi}_{\vec{n}_{2}}(\hat{t}\,)\right\rangle ^{\left(0\right)}=\left\langle \hat{\xi}_{\vec{n}_{1}}\hat{\xi}_{\vec{n}_{2}}\right\rangle ^{\left(0\right)}e^{-|\vec{n}_{2}|^{4}\hat{t}} \,. \label{eq:linear dyn 2pt zeroth order}
\end{equation}
This exponential decay is of course the expected result for a completely linear theory. As with the static structure factor, corrections can be obtained by subbing $\mathbb{F}=\hat{\xi}_{\vec{n}_{1}}\left(0\right)\hat{\xi}_{\vec{n}_{2}}(\hat{t}\,)$ with $m=1$ into equation~(\ref{eq:moment eq nondim linear})
\begin{multline}
\frac{\partial}{\partial\hat{t}}\left\langle \hat{\xi}_{\vec{n}_{1}}\left(0\right)\hat{\xi}_{\vec{n}_{2}}(\hat{t}\,)\right\rangle ^{\left(1\right)}=-\left|\vec{n}_{2}\right|^{4}\left\langle \hat{\xi}_{\vec{n}_{1}}\left(0\right)\hat{\xi}_{\vec{n}_{2}}\right\rangle ^{\left(1\right)}\\-\frac{1}{2}\frac{1}{g^{2}}\sum_{\begin{subarray}{c}
\vec{\ell}_{1}\ne\vec{n}_{2}\\
\vec{\ell}_{2},\vec{\ell}_{3}
\end{subarray}}V_{\vec{n}_{2},\vec{\ell}_{1},\vec{\ell}_{2},\vec{\ell}_{3}}\left\langle \hat{\xi}_{\vec{n}_{1}}\left(0\right)\hat{\xi}_{\vec{\ell}_{1}}\hat{\xi}_{\vec{\ell}_{2}}\hat{\xi}_{\vec{\ell}_{3}}\right\rangle ^{\left(0\right)} \label{eq:linear dyn 2pt func derivation}
\end{multline}
and as before, we find that we need the zeroth order time-dependent four-point function to proceed. Subbing $\mathbb{F}=\hat{\xi}_{\vec{n}_{1}}\left(0\right)\hat{\xi}_{\vec{\ell}_{1}}\hat{\xi}_{\vec{\ell}_{2}}\hat{\xi}_{\vec{\ell}_{3}}$ and $m=0$ into equation~(\ref{eq:moment eq nondim linear}) gives the ODE
\begin{multline}
\frac{\partial}{\partial\hat{t}}\left\langle \hat{\xi}_{\vec{n}_{1}}\left(0\right)\hat{\xi}_{\vec{\ell}_{1}}\hat{\xi}_{\vec{\ell}_{2}}\hat{\xi}_{\vec{\ell}_{3}}\right\rangle ^{\left(0\right)}= 
-\left(\left|\vec{\ell}_{1}\right|^{4}+\left|\vec{\ell}_{2}\right|^{4}+\left|\vec{\ell}_{3}\right|^{4}\right)\left\langle \hat{\xi}_{\vec{n}_{1}}\left(0\right)\hat{\xi}_{\vec{\ell}_{1}}\hat{\xi}_{\vec{\ell}_{2}}\hat{\xi}_{\vec{\ell}_{3}}\right\rangle ^{\left(0\right)} \\
+\delta_{\vec{\ell}_{1},-\vec{\ell}_{2}}\left\langle \hat{\xi}_{\vec{n}_{1}}\left(0\right)\hat{\xi}_{\vec{\ell}_{3}}\right\rangle ^{\left(0\right)}+\delta_{\vec{\ell}_{1},-\vec{\ell}_{3}}\left\langle \hat{\xi}_{\vec{n}_{1}}\left(0\right)\hat{\xi}_{\vec{\ell}_{2}}\right\rangle ^{\left(0\right)} 
+\delta_{\vec{\ell}_{2},-\vec{\ell}_{3}}\left\langle \hat{\xi}_{\vec{n}_{1}}\left(0\right)\hat{\xi}_{\vec{\ell}_{1}}\right\rangle ^{\left(0\right)} \,. \label{eq:lin dyn 4 pt ODE zeroth order}
\end{multline}
Focusing on any of the last three terms of this equation, note that equations~(\ref{eq:linear stat 2pt zeroth order}) and (\ref{eq:linear dyn 2pt zeroth order}) can be used to write
\begin{multline}
\delta_{\vec{\ell}_{1},-\vec{\ell}_{2}}\left\langle \hat{\xi}_{\vec{n}_{1}}\left(0\right)\hat{\xi}_{\vec{\ell}_{3}}\right\rangle ^{\left(0\right)}= \\
\left(\left|\vec{\ell}_{1}\right|^{4}+\left|\vec{\ell}_{2}\right|^{4}+\left|\vec{\ell}_{3}\right|^{4}-\left|\vec{n}_{1}\right|^{4}\right)\left\langle \hat{\xi}_{\vec{\ell}_{1}}\hat{\xi}_{\vec{\ell}_{2}}\right\rangle ^{\left(0\right)}\left\langle \hat{\xi}_{\vec{n}_{1}}\hat{\xi}_{\vec{\ell}_{3}}\right\rangle ^{\left(0\right)}e^{-|\vec{n}_{1}|^{4}\hat{t}} \,,
\end{multline}
where we have added and subtracted $\left|\vec{\ell}_{3}\right|^{4}$ and $\left|\vec{n}_{1}\right|^{4}$ since the two point function $\left\langle \hat{\xi}_{\vec{n}_{1}}\hat{\xi}_{\vec{\ell}_{3}}\right\rangle ^{\left(0\right)}$ anyway vanishes unless $\vec{\ell}_{3}=-\vec{n}_{1}$. Accordingly, upon comparison with equation~(\ref{eq:linear stat 4pt zeroth order}), the sum of the last three terms of equation~(\ref{eq:lin dyn 4 pt ODE zeroth order}) can be written as
\begin{multline}
\delta_{\vec{\ell}_{1},-\vec{\ell}_{2}}\left\langle \hat{\xi}_{\vec{n}_{1}}\left(0\right)\hat{\xi}_{\vec{\ell}_{3}}\right\rangle ^{\left(0\right)}+\delta_{\vec{\ell}_{1},-\vec{\ell}_{3}}\left\langle \hat{\xi}_{\vec{n}_{1}}\left(0\right)\hat{\xi}_{\vec{\ell}_{2}}\right\rangle ^{\left(0\right)} 
+\delta_{\vec{\ell}_{2},-\vec{\ell}_{3}}\left\langle \hat{\xi}_{\vec{n}_{1}}\left(0\right)\hat{\xi}_{\vec{\ell}_{1}}\right\rangle ^{\left(0\right)}= \\
\left(\left|\vec{\ell}_{1}\right|^{4}+\left|\vec{\ell}_{2}\right|^{4}+\left|\vec{\ell}_{3}\right|^{4}-\left|\vec{n}_{1}\right|^{4}\right)\left\langle \hat{\xi}_{\vec{n}_{1}}\hat{\xi}_{\vec{\ell}_{1}}\hat{\xi}_{\vec{\ell}_{2}}\hat{\xi}_{\vec{\ell}_{3}}\right\rangle ^{\left(0\right)}e^{-|\vec{n}_{1}|^{4}\hat{t}} \,,
\end{multline}
and thus it becomes easy to check that the solution to equation (\ref{eq:lin dyn 4 pt ODE zeroth order}) is unsurprisingly
\begin{equation}
\left\langle \hat{\xi}_{\vec{n}_{1}}\left(0\right)\hat{\xi}_{\vec{\ell}_{1}}\hat{\xi}_{\vec{\ell}_{2}}\hat{\xi}_{\vec{\ell}_{3}}\right\rangle ^{\left(0\right)}=\left\langle \hat{\xi}_{\vec{n}_{1}}\hat{\xi}_{\vec{\ell}_{1}}\hat{\xi}_{\vec{\ell}_{2}}\hat{\xi}_{\vec{\ell}_{3}}\right\rangle ^{\left(0\right)}e^{-|\vec{n}_{1}|^{4}\hat{t}} \,.
\end{equation}
Returning to equation~(\ref{eq:linear dyn 2pt func derivation}) and making use of equation~(\ref{eq:V sum}), we obtain the following nonhomogeneous ODE for the first order time-dependent two-point function
\begin{equation}
\frac{\partial}{\partial\hat{t}}\left\langle \hat{\xi}_{\vec{n}_{1}}\left(0\right)\hat{\xi}_{\vec{n}_{2}}(\hat{t}\,)\right\rangle ^{\left(1\right)}=-\left|\vec{n}_{2}\right|^{4}\left\langle \hat{\xi}_{\vec{n}_{1}}\left(0\right)\hat{\xi}_{\vec{n}_{2}}(\hat{t}\,)\right\rangle ^{\left(1\right)}\\-\frac{3\pi}{8g^{2}}\left|\vec{n}_{2}\right|^{2}\left\langle \hat{\xi}_{\vec{n}_{1}}\hat{\xi}_{\vec{n}_{2}}\right\rangle ^{\left(0\right)}e^{-|\vec{n}_{2}|^{4}\hat{t}} \,. \label{eq:1st order linear dyn 2pt ODE}
\end{equation}
Here, we have freely replaced the decay rate of the exponential $|\vec{n}_1|^4$ with $|\vec{n}_2|^4$ since the static two-point function ensures that this term vanishes anyway if $\vec{n}_1\ne -\vec{n}_2$. We have also replaced the sum, now given by equation~(\ref{eq:V sum}), with its approximate value given by equation~(\ref{eq:evaluated integral}).

Equation~(\ref{eq:1st order linear dyn 2pt ODE}) is easily solved but the fact that its nonhomogeneous part decays with the decay rate of the homogeneous part results in the solution containing a secular term
\begin{equation}
\left\langle \hat{\xi}_{\vec{n}_{1}}\left(0\right)\hat{\xi}_{\vec{n}_{2}}(\hat{t}\,)\right\rangle ^{\left(1\right)}=\\
\left[\left\langle \hat{\xi}_{\vec{n}_{1}}\hat{\xi}_{\vec{n}_{2}}\right\rangle ^{\left(1\right)}-\frac{3\pi}{8g^{2}}\left|\vec{n}_{2}\right|^{2}\left\langle \hat{\xi}_{\vec{n}_{1}}\hat{\xi}_{\vec{n}_{2}}\right\rangle ^{\left(0\right)}\hat{t}\,\right]e^{-|\vec{n}_{2}|^{4}\hat{t}} \,. \label{eq:linear dyn 2pt secular}
\end{equation}
This secular term is an unphysical artefact of the perturbation theory which limits the solution to short times. If we wish to obtain a first order correction which is valid for all times, we need to perform a more sophisticated perturbation theory, such as the Poincar\'e-Lindstedt method \cite{Poincare1893, Lindstedt1882, DrazinBook}, in which the decay rate of equation~(\ref{eq:moment eq nondim linear}) is also expanded order by order. While such an analysis is possible, it is faster to simply note that equation~(\ref{eq:linear dyn 2pt secular}) is merely the large $g$ expansion of
\begin{equation}
\left\langle \hat{\xi}_{\vec{n}_{1}}\left(0\right)\hat{\xi}_{\vec{n}_{2}}(\hat{t}\,)\right\rangle ^{\left(1\right)}=\\
\left\langle \hat{\xi}_{\vec{n}_{1}}\hat{\xi}_{\vec{n}_{2}}\right\rangle ^{\left(1\right)}e^{-\left|\vec{n}_{2}\right|^{4}\left(1+\frac{3\pi}{8g^{2}}\frac{1}{\left|\vec{n}_{2}\right|^{2}}\right)\hat{t}}
\end{equation}
where we have used in the exponential the renormalised linear coupling implied by equation~(\ref{eq:pade linear}). Accordingly, we have found that in the linear regime, the weakly coupled nonlinearity merely has the effect of renormalising the linear coupling $n^4\rightarrow n^{4}\left(1+\frac{3\pi}{8g^{2}}\frac{1}{n^{2}}\right)$, both for static and dynamic quantities.

In dimensional terms, we obtain
\begin{equation}
\left\langle \tilde{\xi}_{\vec{n}_{1}}\left(0\right)\tilde{\xi}_{\vec{n}_{2}}(t\,)\right\rangle ^{\left(1\right)}=\\\left\langle \tilde{\xi}_{\vec{n}_{1}}\tilde{\xi}_{\vec{n}_{2}}\right\rangle ^{\left(1\right)}e^{-\left(2\pi\right)^{4}\frac{\kappa\left|\vec{n}_{2}\right|^{4}}{\alpha L^{4}}\left(1+\frac{3\pi}{8g^{2}}\frac{1}{\left|\vec{n}_{2}\right|^{2}}\right)t} \,. \label{eq:lin dyn 2 pt dimensional}
\end{equation}

In Fourier space, the dynamical scaling described by equation~(\ref{eq:exponent definitions}) takes the form \cite{Katzav2004}
\begin{equation}
\left\langle \tilde{\xi}_{\vec{n}}\left(0\right)\tilde{\xi}_{-\vec{n}}\left(t\right)\right\rangle =\left\langle \tilde{\xi}_{\vec{n}}\tilde{\xi}_{-\vec{n}}\right\rangle \Phi\left(n^{z}t\right)
\end{equation}
where $z$ denotes the dynamic exponent and $\Phi(u)$ denotes a scaling function. Direct comparison with equation~(\ref{eq:lin dyn 2 pt dimensional}) immediately shows that for our system, $z=4$ and $\Phi(u)$ exhibits an exponentially decay.

Indeed, for any linear system, the roughness exponent $\zeta$ and dynamic exponent $z$ will be related by
\begin{equation}
z = 2\zeta + d \,, \label{eq:exponent relationship}
\end{equation}
where $d$ denotes the dimension of the system \cite{Katzav2011b, Katzav2011a}. Accordingly, for our case of $d=2$, it is unsurprising to find that the values $\zeta = 1$ and $z = 4$ satisfy this relationship. Importantly however, the fact that this linear character extends beyond the large-scale behaviour described by the exponents to subsequent corrections is nontrivial. Indeed, in the next section, we will unexpectedly find that this linear character over all scales remains even in the context of strong nonlinear coupling and is not merely an artefact of weak coupling.

\section{The Nonlinear Regime\label{sec:nonlinear theory}}

While ordinary everyday systems are expected to reside in the linear regime of $g \gg 1$, it is not difficult to imagine circumstances where $g \ll 1$ and thus the nonlinear term will dominate equation~(\ref{eq:moment eq nondim}). In particular, since equation~(\ref{eq:g definition 2}) shows that $g\sim h^{5/2}$, we expect $g$ to be small for sufficiently thin materials such as graphene or biological membranes. Indeed, single layer graphene will have the thickness of a single atom, i.e. $h\sim1\textrm{ \AA}$, and thus a sheet of size $L\sim250\textrm{ nm}$ \cite{Rao2009} with Young's modulus $E\sim2\textrm{ TPa}$ \cite{Lee2012} and bending rigidity $\kappa\sim2\times10^{-19}\textrm{ N m}$ \cite{Wei2013} at temperatures of say $T\sim300\textrm{ K}$ will have $g\sim 0.02$ which is indeed much smaller than unity. Similarly, polymerised lipid bilayers surrounding cells or organelles ($L\sim 1\textrm{ \si{\micro\meter}}-100\textrm{ \si{\micro\meter}}$) with thickness $h\sim 5\textrm{ nm}$ and Young's modulus $E\sim 20\textrm{ MPa}$ \cite{Picas2012} will have values of  $g$ ranging from $\sim 0.003 $ to $\sim 0.3 $. A theory which can also describe the small $g$ nonlinear regime is thus certainly of value.

To investigate this regime, it will be helpful to rescale equation~(\ref{eq:moment eq nondim}) as follows. If we define a new dimensionless time $\bar{t}$ and vertical displacement $\bar{\xi}_{\vec{n}}$ as
\begin{equation}
\bar{t}=\frac{1}{g}\hat{t}=\left[\frac{\left(2\pi\right)^{4}2k_{\mathrm{B}}TEh}{\alpha^{2}L^{6}}\right]^{1/2}t \label{eq:t scaling}
\end{equation}
and
\begin{equation}
\bar{\xi}_{\vec{n}}(\bar{t}\,)=\frac{1}{g^{1/2}}\hat{\xi}_{\vec{n}}(\hat{t}\,) =\left[\frac{\left(2\pi\right)^{4}Eh}{2k_{\mathrm{B}}TL^{2}}\right]^{1/4}\tilde{\xi}_{\vec{n}}\left(t\right) \,, \label{eq:xi scaling}
\end{equation}
then equation~(\ref{eq:moment eq nondim}) can be rewritten as
\begin{multline}
\frac{\partial\left\langle \mathbb{F}\right\rangle }{\partial\bar{t}}=\frac{1}{2}\sum_{\vec{n}}\left\langle \frac{\partial^{2}\mathbb{F}}{\partial\bar{\xi}_{\vec{n}}\partial\bar{\xi}_{-\vec{n}}}\right\rangle -g\sum_{\vec{n}}\left|\vec{n}\right|^{4}\left\langle \frac{\partial\mathbb{F}}{\partial\bar{\xi}_{\vec{n}}}\bar{\xi}_{\vec{n}}\right\rangle \\-\frac{1}{2}\sum_{\vec{n}}\sum_{\begin{subarray}{c}
\vec{\ell}_{1}\ne\vec{n}\\
\vec{\ell}_{2},\vec{\ell}_{3}
\end{subarray}}V_{\vec{n},\vec{\ell}_{1},\vec{\ell}_{2},\vec{\ell}_{3}}\left\langle \frac{\partial\mathbb{F}}{\partial\bar{\xi}_{\vec{n}}}\bar{\xi}_{\vec{\ell}_{1}}\bar{\xi}_{\vec{\ell}_{2}}\bar{\xi}_{\vec{\ell}_{3}}\right\rangle \,, \label{eq:moment eq nondim nonlinear}
\end{multline}
from which it is clear that $g$ is coupled to the linear part. Accordingly, for $g\ll 1$ attempting to perturb the equation around the linear part is singular and thus more sophisticated techniques are required if we wish to understand this regime.

The Self-Consistent Expansion (SCE) is such a technique. Originally developed to investigate the KPZ equation \cite{Schwartz1992, Schwartz1998, Katzav1999, Katzav2002, Schwartz2002, Katzav2002a, Katzav2003, Katzav2004, Katzav2004a}, the SCE has been successfully used to study a variety of out-of-equilibrium systems including turbulence, fracture and wetting fronts \cite{Edwards2002, Katzav2006, Katzav2007}. More recently it was shown that the SCE is able to produce excellent results at low order for the zero-dimensional $\phi^4$-theory and even exact convergence at high orders \cite{Schwartz2008, Remez2018}. On this basis, the SCE was used in \cite{Steinbock2022} and \cite{Steinbock2023} to study the related problem of a sheet fluctuating under conserved noise immediately prior to crumpling and following this success, here too we will apply the SCE to equation~(\ref{eq:moment eq nondim nonlinear}).

The idea behind the SCE is that we can expand equation~(\ref{eq:moment eq nondim nonlinear}) around any linear theory, not just the one ``handed to us''. That is, consider some arbitrary linear coupling $\Gamma_{|\vec{n}|}$. We can identically insert this coupling into equation~(\ref{eq:moment eq nondim nonlinear}) by writing
\begin{multline}
\frac{\partial\left\langle \mathbb{F}\right\rangle }{\partial\bar{t}}=\frac{1}{2}\sum_{\vec{n}}\left\langle \frac{\partial^{2}\mathbb{F}}{\partial\bar{\xi}_{\vec{n}}\partial\bar{\xi}_{-\vec{n}}}\right\rangle-\sum_{\vec{n}}\Gamma_{|\vec{n}|}\left\langle \frac{\partial\mathbb{F}}{\partial\bar{\xi}_{\vec{n}}}\bar{\xi}_{\vec{n}}\right\rangle
\\ -\sum_{\vec{n}}\left(g\left|\vec{n}\right|^{4}-\Gamma_{|\vec{n}|}\right)\left\langle \frac{\partial\mathbb{F}}{\partial\bar{\xi}_{\vec{n}}}\bar{\xi}_{\vec{n}}\right\rangle-\frac{1}{2}\sum_{\vec{n}}\sum_{\begin{subarray}{c}
\vec{\ell}_{1}\ne\vec{n}\\
\vec{\ell}_{2},\vec{\ell}_{3}
\end{subarray}}V_{\vec{n},\vec{\ell}_{1},\vec{\ell}_{2},\vec{\ell}_{3}}\left\langle \frac{\partial\mathbb{F}}{\partial\bar{\xi}_{\vec{n}}}\bar{\xi}_{\vec{\ell}_{1}}\bar{\xi}_{\vec{\ell}_{2}}\bar{\xi}_{\vec{\ell}_{3}}\right\rangle \,.
\end{multline}

The goal is to choose the linear coupling $\Gamma_{|\vec{n}|}$ such that the expansion around this linear theory is close to the behaviour of the nonlinear system though this task will be deferred to later. For now, we can simply perform an ordinary perturbative expansion in which the nonlinear term and ``residual'' linear term are corrections, that is, we can write
\begin{multline}
\frac{\partial\left\langle \mathbb{F}\right\rangle ^{\left(m\right)}}{\partial\bar{t}}=\frac{1}{2}\sum_{\vec{n}}\left\langle \frac{\partial^{2}\mathbb{F}}{\partial\bar{\xi}_{\vec{n}}\partial\bar{\xi}_{-\vec{n}}}\right\rangle ^{\left(m\right)}-\sum_{\vec{n}}\Gamma_{|\vec{n}|}\left\langle \frac{\partial\mathbb{F}}{\partial\bar{\xi}_{\vec{n}}}\bar{\xi}_{\vec{n}}\right\rangle ^{\left(m\right)}\\-\sum_{\vec{n}}\left(g\left|\vec{n}\right|^{4}-\Gamma_{|\vec{n}|}\right)\left\langle \frac{\partial\mathbb{F}}{\partial\bar{\xi}_{\vec{n}}}\bar{\xi}_{\vec{n}}\right\rangle ^{\left(m-1\right)}-\frac{1}{2}\sum_{\vec{n}}\sum_{\begin{subarray}{c}
\vec{\ell}_{1}\ne\vec{n}\\
\vec{\ell}_{2},\vec{\ell}_{3}
\end{subarray}}V_{\vec{n},\vec{\ell}_{1},\vec{\ell}_{2},\vec{\ell}_{3}}\left\langle \frac{\partial\mathbb{F}}{\partial\bar{\xi}_{\vec{n}}}\bar{\xi}_{\vec{\ell}_{1}}\bar{\xi}_{\vec{\ell}_{2}}\bar{\xi}_{\vec{\ell}_{3}}\right\rangle ^{\left(m-1\right)} \,. \label{eq:SCE}
\end{multline}

In the following subsections, we will analyse the static and dynamic structure factors which emerge from such a perturbative expansion.

\subsection{The Static Structure Factor\label{sec:nonlinear static structure factor}}

As with the linear regime, the equal-time two-point and four-point functions can be obtained by subbing in $\mathbb{F}=\bar{\xi}_{\vec{n}_{1}}\bar{\xi}_{\vec{n}_{2}}$ and $\mathbb{F}=\bar{\xi}_{\vec{n}_{1}}\bar{\xi}_{\vec{\ell}_{1}}\bar{\xi}_{\vec{\ell}_{2}}\bar{\xi}_{\vec{\ell}_{3}}$ and similarly to before, at zeroth order, we obtain
\begin{equation}
\left\langle \bar{\xi}_{\vec{n}_{1}}\bar{\xi}_{\vec{n}_{2}}\right\rangle ^{\left(0\right)}=\frac{\delta_{\vec{n}_{1},-\vec{n}_{2}}}{\Gamma_{|\vec{n}_{1}|}+\Gamma_{|\vec{n}_{2}|}}=\frac{\delta_{\vec{n}_{1},-\vec{n}_{2}}}{2\Gamma_{|\vec{n}_{1}|}} \label{eq:nonlin static 2 pt zeroth order}
\end{equation}
and Isserlis' theorem
\begin{multline}
\left\langle \bar{\xi}_{\vec{n}_{1}}\bar{\xi}_{\vec{\ell}_{1}}\bar{\xi}_{\vec{\ell}_{2}}\bar{\xi}_{\vec{\ell}_{3}}\right\rangle ^{\left(0\right)}=\left\langle \bar{\xi}_{\vec{n}_{1}}\bar{\xi}_{\vec{\ell}_{1}}\right\rangle ^{\left(0\right)}\left\langle \bar{\xi}_{\vec{\ell}_{2}}\bar{\xi}_{\vec{\ell}_{3}}\right\rangle ^{\left(0\right)}\\+\left\langle \bar{\xi}_{\vec{n}_{1}}\bar{\xi}_{\vec{\ell}_{2}}\right\rangle ^{\left(0\right)}\left\langle \bar{\xi}_{\vec{\ell}_{1}}\bar{\xi}_{\vec{\ell}_{3}}\right\rangle ^{\left(0\right)}+\left\langle \bar{\xi}_{\vec{n}_{1}}\bar{\xi}_{\vec{\ell}_{3}}\right\rangle ^{\left(0\right)}\left\langle \bar{\xi}_{\vec{\ell}_{1}}\bar{\xi}_{\vec{\ell}_{2}}\right\rangle ^{\left(0\right)} \,. \label{eq:isserlis 2}
\end{multline}
Proceeding to first order, subbing $\mathbb{F}=\bar{\xi}_{\vec{n}_{1}}\bar{\xi}_{\vec{n}_{2}}$ and $m=1$ into equation~(\ref{eq:SCE}) gives 
\begin{multline}
0=\delta_{\vec{n}_{1},-\vec{n}_{2}}-\left(\Gamma_{|\vec{n}_{1}|}+\Gamma_{|\vec{n}_{2}|}\right)\left\langle \bar{\xi}_{\vec{n}_{1}}\bar{\xi}_{\vec{n}_{2}}\right\rangle ^{\left(1\right)}\\-\left[g\left(\left|\vec{n}_{1}\right|^{4}+\left|\vec{n}_{2}\right|^{4}\right)-\left(\Gamma_{|\vec{n}_{1}|}+\Gamma_{|\vec{n}_{2}|}\right)\right]\left\langle \bar{\xi}_{\vec{n}_{1}}\bar{\xi}_{\vec{n}_{2}}\right\rangle ^{\left(0\right)}\\-\frac{1}{2}\sum_{\begin{subarray}{c}
\vec{\ell}_{1}\ne\vec{n}_{1}\\
\vec{\ell}_{2},\vec{\ell}_{3}
\end{subarray}}\left[V_{\vec{n}_{1},\vec{\ell}_{1},\vec{\ell}_{2},\vec{\ell}_{3}}\left\langle \bar{\xi}_{\vec{n}_{2}}\bar{\xi}_{\vec{\ell}_{1}}\bar{\xi}_{\vec{\ell}_{2}}\bar{\xi}_{\vec{\ell}_{3}}\right\rangle ^{\left(0\right)}+V_{\vec{n}_{2},\vec{\ell}_{1},\vec{\ell}_{2},\vec{\ell}_{3}}\left\langle \bar{\xi}_{\vec{n}_{1}}\bar{\xi}_{\vec{\ell}_{1}}\bar{\xi}_{\vec{\ell}_{2}}\bar{\xi}_{\vec{\ell}_{3}}\right\rangle ^{\left(0\right)} \right] \,,
\end{multline}
which after simplifying the sums by subbing in equation~(\ref{eq:isserlis 2}) and rearranging gives
\begin{multline}
\left\langle \bar{\xi}_{\vec{n}_{1}}\bar{\xi}_{\vec{n}_{2}}\right\rangle ^{\left(1\right)}=\left\langle \bar{\xi}_{\vec{n}_{1}}\bar{\xi}_{\vec{n}_{2}}\right\rangle ^{\left(0\right)} \\
+\frac{\left\langle \bar{\xi}_{\vec{n}_{1}}\bar{\xi}_{\vec{n}_{2}}\right\rangle ^{\left(0\right)}}{\Gamma_{|\vec{n}_{1}|}}\left(\Gamma_{|\vec{n}_{1}|}-g\left|\vec{n}_{1}\right|^{4}-\frac{1}{2}\sum_{\vec{\ell}\ne\vec{n}_{1}}\frac{\left|\vec{n}_{1}\times\vec{\ell}\,\right|^{4}}{\Gamma_{\left|\vec{\ell}\,\right|}\left|\vec{n}_{1}-\vec{\ell}\,\right|^{4}}\right) \,.
\end{multline}

We now argue that the appropriate linear theory to expand around is the one for which subsequent corrections to the linear theory are minimised, that is, we self-consistently desire that
\begin{equation}
\left\langle \bar{\xi}_{\vec{n}_{1}}\bar{\xi}_{\vec{n}_{2}}\right\rangle ^{\left(1\right)}=\left\langle \bar{\xi}_{\vec{n}_{1}}\bar{\xi}_{\vec{n}_{2}}\right\rangle ^{\left(0\right)} \,.
\end{equation}

Clearly this will be the case if $\Gamma_{|\vec{n}|}$ obeys the discrete integral equation
\begin{equation}
\Gamma_{|\vec{n}|}=g\left|\vec{n}\right|^{4}+\frac{1}{2}\sum_{\vec{\ell}\ne\vec{n}}\frac{\left|\vec{n}\times\vec{\ell}\,\right|^{4}}{\Gamma_{\left|\vec{\ell}\,\right|}\left|\vec{n}-\vec{\ell}\,\right|^{4}} \,. \label{eq:disc integ eq}
\end{equation}
This equation defines the linear coupling which we have desired to expand around and though such equations are generically difficult to solve (indeed, there may not exist an exact solution to this equation), we can relatively easily obtain corresponding approximate asymptotic behaviours of $\Gamma_{|\vec{n}|}$ by considering the limit of a continuous spectrum, though this may result in missing solutions which also contain a dependence on the direction of $\vec{n}$. The discrete frequencies $\vec{n}$ of our system will appear continuous in the limit where the size of the sheet $L$ is taken to infinity thus it is natural to define the scaled frequencies $\vec{q} = \vec{n}/L$ and $\vec{k} = \vec{\ell}/L$. At the same time, it will be helpful to scale $\Gamma_{|\vec{n}|}$ by some amount to keep it finite and so we define $\Gamma(q)=\Gamma_{|\vec{n}|}/L^\nu$ where $\nu$ is an appropriate scaling exponent which will become clear shortly. Then equation~(\ref{eq:disc integ eq}) becomes
\begin{equation}
\Gamma\left(q\right)=L^{4-\nu}gq^{4}+\frac{1}{2}L^{6-2\nu}\sum_{\vec{k}\ne\vec{q}}\frac{1}{L^{2}}\frac{\left|\vec{q}\times\vec{k}\,\right|^{4}}{\Gamma\left(k\right)\left|\vec{q}-\vec{k}\,\right|^{4}} \,.
\end{equation}

Setting $\nu = 3$ and taking the limit $L\rightarrow\infty$, we obtain the continuous integral equation for $\Gamma(q)$
\begin{equation}
\Gamma\left(q\right)=(Lg)q^{4}+\frac{1}{2}\int d^{2}k\frac{\left|\vec{q}\times\vec{k}\,\right|^{4}}{\Gamma\left(k\right)\left|\vec{q}-\vec{k}\,\right|^{4}} \,.
\end{equation}
It is worth noting that the coefficient $Lg$ remains constant as $L\rightarrow\infty$ since $g\sim 1/L$ and thus this continuous integral equation is a well-defined continuous analogue of our discrete integral equation. We can write the integral in polar coordinates as
\begin{equation}
\Gamma\left(q\right)=(Lg)q^{4} \\
+\frac{q^{4}}{2}\int_{0}^{q_\mathrm{max}}dk\,\frac{k^{5}}{\Gamma\left(k\right)}\int_{0}^{2\pi}d\theta\,\frac{\sin^{4}\theta}{\left(q^{2}+k^{2}-2qk\cos\theta\right)^{2}} \,,
\end{equation}
where $q_\mathrm{max}=n_\mathrm{max}/L$ denotes a potential upper frequency cut-off. As mentioned towards the end of Section~\ref{sec:linear static structure factor}, if the system has some finite resolution scale $a\ll L$, an upper frequency cut-off $n_\mathrm{max}=L/(2a)\gg 1$ is introduced. It will become clear that in the nonlinear regime, the presence of a finite upper frequency cut-off can result in very different behaviour and thus we already allow for such a possibility at this stage. The case with no upper cut-off can always be obtained later by simply taking the limit $q_\mathrm{max}\rightarrow\infty$.

The angular integral has already been encountered and is given by equation~(\ref{eq:angular integral}) thus we can obtain the following integral equation for $\Gamma(q)$
\begin{equation}
\Gamma\left(q\right)=(Lg)q^{4} \\
+\frac{3\pi}{8}\left[\int_{0}^{q}dk\,\frac{k^{5}}{\Gamma\left(k\right)}+q^{4}\int_{q}^{q_{\mathrm{max}}}dk\,\frac{k}{\Gamma\left(k\right)}\right] \label{eq:cont integ eq}
\end{equation}
and with this equation in hand, we are ready to determine the asymptotic behaviours of $\Gamma(q)$. 

Beginning with large $q$, suppose that $\Gamma(q)$ has an asymptotic expansion
\begin{equation}
\Gamma(q)\sim A_\infty q^{\nu_\infty } \,, \label{eq:large q ansatz}
\end{equation}
where $A_\infty $ and $\nu_\infty $ determine the leading order behaviour of $\Gamma(q)$ and higher order corrections are neglected. We can write equation~(\ref{eq:cont integ eq}) as
\begin{equation}
\Gamma\left(q\right)=(Lg)q^{4}+\frac{3\pi}{8}\int_{q}^{q_{\mathrm{max}}}dk\,\frac{k\left(q^{4}-k^{4}\right)}{\Gamma\left(k\right)}+C_\infty  \,, \label{eq:integ eq large q}
\end{equation}
where
\begin{equation}
C_\infty =\frac{3\pi}{8}\int_{0}^{q_{\mathrm{max}}}dk\,\frac{k^{5}}{\Gamma\left(k\right)}
\end{equation}
denotes a constant. Now subbing equation~(\ref{eq:large q ansatz}) into equation~(\ref{eq:integ eq large q}), we obtain at leading order that
\begin{multline}
A_\infty q^{\nu_\infty }=\left[(Lg)-\frac{3\pi}{8A_\infty }\frac{q_{\mathrm{max}}^{2-\nu_\infty }}{\left(\nu_\infty -2\right)}\right]q^{4}+\frac{3\pi}{8A_\infty }\frac{4}{\left(6-\nu_\infty \right)\left(\nu_\infty -2\right)}q^{6-\nu_\infty }\\+C_\infty -\frac{3\pi}{8A_\infty ^{\left(0\right)}}\frac{q_{\mathrm{max}}^{6-\nu_\infty }}{\left(6-\nu_\infty \right)} \,, \label{eq:large q asymptotic}
\end{multline}
where we have implicitly assumed that $\nu_\infty \ne 2,6$. (Assuming $\nu_\infty =2,6$ results in logarithmic corrections on the right-hand side which cannot be balanced and thus this assumption is justified.) Examination of this equation reveals that there are two choices for $\nu_\infty $ which can satisfy the equation at leading order, either $\nu_\infty  = 4$ or $\nu_\infty  = 3$. If $\nu_\infty  = 4$, then we must also have
\begin{equation}
A_\infty =(Lg)-\frac{3\pi}{8A_\infty }\frac{1}{2q_{\mathrm{max}}^{2}} \,,
\end{equation}
which can be solved for $A_\infty $ to give
\begin{equation}
A_\infty =\frac{(Lg)}{2}\left[1+\sqrt{1-\frac{3\pi}{4}\frac{1}{(Lgq_{\mathrm{max}})^{2}}}\right] \,. \label{eq:A>}
\end{equation}
We thus find that the first option for the large $q$ asymptotic behaviour of $\Gamma(q)$ is 
\begin{equation}
\Gamma(q)\sim\frac{(Lg)}{2}\left[1+\sqrt{1-\frac{3\pi}{4}\frac{1}{(Lgq_{\mathrm{max}})^{2}}}\right]q^{4} \,.
\end{equation}
Recalling that $\Gamma(q)=\Gamma_{|\vec{n}|}/L^3$ and $q = n/L$, we find that this asymptotic behaviour corresponds to 
\begin{equation}
\Gamma_{\left|\vec{n}\right|}\sim\frac{g}{2}\left[1+\sqrt{1-\frac{3\pi}{4}\frac{1}{(gn_{\mathrm{max}})^{2}}}\right]n^{4} \,,
\end{equation}
or when $gn_{\mathrm{max}}\gg1$
\begin{equation}
\Gamma_{\left|\vec{n}\right|}\sim gn^{4}\left[1-\frac{3\pi}{16}\frac{1}{(gn_{\mathrm{max}})^{2}}\right] \,.
\end{equation}
Noting that
\begin{equation}
\left\langle \hat{\xi}_{\vec{n}}\hat{\xi}_{-\vec{n}}\right\rangle ^{\left(0\right)}=g\left\langle \bar{\xi}_{\vec{n}}\bar{\xi}_{-\vec{n}}\right\rangle ^{\left(0\right)}=\frac{g}{2\Gamma_{|\vec{n}|}} \,,
\end{equation}
and comparing with equation~(\ref{eq:lin stat 2 pt w cutoff}), we thus find that at leading order, this large frequency asymptotic behaviour completely agrees with the analysis performed for the linear regime. It is important however to point out that agreement merely requires that $gn_{\mathrm{max}}\gg1$ rather than $g\gg1$ and thus if $n_{\mathrm{max}}$ is sufficiently large, it is possible to obtain a linear-like response for large frequencies, even if $g\ll 1$. If no upper frequency cut-off exists whatsoever, this large frequency linear-like response is actually obtained for any nonzero $g$.

At the same time, from equation~(\ref{eq:A>}), it is apparent that if 
\begin{equation}
gn_{\mathrm{max}}=Lgq_{\mathrm{max}}<\sqrt{\frac{3\pi}{4}}\approx1.53... \,,
\end{equation}
then $A_\infty $ will cease to be real, i.e. $A_\infty  \in \mathbb{C}$, which cannot be the case. Accordingly, when $gn_\mathrm{max}$ is sufficiently small, we find that the linear $n^4$ asymptotic behaviour becomes untenable. Under these circumstances, the other possible asymptotic behaviour mentioned above emerges, i.e. we must have $\nu_\infty  = 3$ in which case we simply find that
\begin{equation}
A_\infty  =\sqrt{\frac{\pi}{2}} \,.
\end{equation}
This eventuality however can only be realised if the coefficient of the $q^4$ term in equation~(\ref{eq:large q asymptotic}) vanishes, that is we must also have
\begin{equation}
Lgq_{\mathrm{max}}=\frac{3\pi}{8A_\infty } \,,
\end{equation}
i.e.
\begin{equation}
gn_{\mathrm{max}}\approx\frac{3\sqrt{2\pi}}{8}\approx0.94... \,.
\end{equation}
We thus find that for values of $gn_{\mathrm{max}}$ close to or smaller than 1, the linear-like quartic large frequency behaviour ceases to exist and a cubic behaviour becomes dominant. 

Moving onto the small $q$ behaviour, suppose that $\Gamma(q)$ has a small $q$ expansion
\begin{equation}
\Gamma(q)\sim A_0 q^{\nu_0 } \,, \label{eq:small q ansatz}
\end{equation}
where $A_\infty $ and $\nu_0 $ determine the small $q$ leading order behaviour. We can write equation~(\ref{eq:cont integ eq}) as
\begin{equation}
\Gamma\left(q\right)=\left[(Lg)+C_0 \right]q^{4}\\
+\frac{3\pi}{8}\left[\int_{0}^{q}dk\,\frac{k^{5}}{\Gamma\left(k\right)}-q^{4}\int_{\lambda}^{q}dk\,\frac{k}{\Gamma\left(k\right)}\right] \,, \label{eq:small q integ equ}
\end{equation}
where
\begin{equation}
C_0 =\frac{3\pi}{8}\int_{\lambda}^{q_{\mathrm{max}}}dk\,\frac{k}{\Gamma\left(k\right)}
\end{equation}
and $\lambda < q$ denotes some small frequency which can be taken to 0 but is temporarily needed to prevent the integrals diverging. Subbing equation~(\ref{eq:small q ansatz}) into equation~(\ref{eq:small q integ equ}) and expanding order by order, we obtain
\begin{equation}
A_0 q^{\nu_0 }=\left[(Lg)+C_0 -\frac{3\pi}{8A_0 }\frac{\lambda^{2-\nu_0 }}{\nu_0 -2}\right]q^{4}\\+\frac{3\pi}{8A_0 }\frac{4}{\left(6-\nu_0 \right)\left(\nu_0 -2\right)}q^{6-\nu_0 } \,,
\end{equation}
which can \textit{only} be satisfied at lowest order by $\nu_0  = 3$, in which case
\begin{equation}
A_0  = \sqrt{\frac{\pi}{2}} \,.
\end{equation}
This leading order cubic small frequency behaviour is obtained independent of the value of $g$ and it is curious to observe that at the critical value $gn_\mathrm{max}=3\sqrt{2\pi}/8$ that $A_0  = A_\infty $. Indeed substituting
\begin{equation}
\Gamma(q)=\sqrt{\frac{\pi}{2}}q^{3}
\end{equation}
directly into equation~(\ref{eq:cont integ eq}) makes it clear that this is the \textit{exact} solution to the integral equation when $ Lgq_\mathrm{max} = gn_\mathrm{max} = 3\sqrt{2\pi}/8 $.

To summarise, we have identified the following three asymptotic behaviours of $\Gamma(q)$
\begin{itemize}
\item For large $q$, when $Lgq_\mathrm{max}\gg 1$, we have found that
\begin{equation}
\Gamma(q)\sim\frac{(Lg)}{2}q^{4} \,.
\end{equation}
This behaviour perfectly corresponds to the behaviour we found in the linear regime.
\item For large $q$, when $Lgq_\mathrm{max} \lesssim 1$, we have found that
\begin{equation}
\Gamma(q)\sim\sqrt{\frac{\pi}{2}}q^{3} \,.
\end{equation}
\item For small $q$, regardless of the value of $g$, we have found that
\begin{equation}
\Gamma(q)\sim\sqrt{\frac{\pi}{2}}q^{3} \,.
\end{equation}
\end{itemize}

Together, these asymptotics suggest that the overall behaviour can be approximated by the combination
\begin{equation}
\Gamma(q)\approx C\left(Lg,q_{\mathrm{max}}\right)q^{3}+A\left(Lg,q_{\mathrm{max}}\right)q^{4} \label{eq:Gamma q final}
\end{equation}
where $A(Lg,q_\mathrm{max})$ and $C\left(Lg,q_{\mathrm{max}}\right)$ denote constants which depend on $Lg$ and $q_\mathrm{max}$, though $C\sim\sqrt{\pi/2}\sim1$ is expected to have a much weaker dependence on $g$ and thus will generally be treated as constant. In terms of the discrete frequencies we began with, this translates to the functional form
\begin{equation}
\Gamma_{\left|\vec{n}\right|}\approx Cn^{3}+A\left(g,n_{\mathrm{max}}\right)n^{4} \,. \label{eq:Gamma n final}
\end{equation}
We now note that the crossover between the $n^3$ and $n^4$ behaviours occurs when $\left|C\right|n^{3}\sim\left|A\left(g,n_{\mathrm{max}}\right)\right|n^{4}$ and thus we can define a crossover frequency $n_c$ as
\begin{equation}
n_{c}\sim\left|\frac{C}{A\left(g,n_{\mathrm{max}}\right)}\right| \,. \label{eq:crossover def}
\end{equation}
At the same time, $n_c$ must obey $1\le n_c \le n_\mathrm{max}$ thus we expect that if $n_c$ is too small, a purely quartic behaviour will be obtained (as was predicted for the linear regime) while if $n_c$ is too large, a purely cubic behaviour will be obtained. For values of $g$ and $n_\mathrm{max}$ which cause $n_c$ to fall between these two extremes, a state of coexistence is obtained in which the cubic behaviour dominates the small frequencies while the quartic behaviour dominates the large frequencies. 

We finish this analysis by providing a useful heuristic for $A(Lg,q_\mathrm{max})$. Differentiating equation~(\ref{eq:cont integ eq}) with respect to $q$, one obtains
\begin{equation}
\Gamma'\left(q\right)=4(Lg)q^{3}+\frac{3\pi}{2}q^{3}\int_{q}^{q_{\mathrm{max}}}dk\,\frac{k}{\Gamma\left(k\right)} \,.
\end{equation}
Subbing in $q = q_\mathrm{max}$ thus gives
\begin{equation}
\Gamma'\left(q_{\mathrm{max}}\right)=4(Lg)q_{\mathrm{max}}^{3} \,.
\end{equation}
At the same time, differentiating equation~(\ref{eq:Gamma q final}) and subbing in $q = q_\mathrm{max}$ gives
\begin{equation}
\Gamma'(q_{\mathrm{max}})\approx3Cq_{\mathrm{max}}^{2}+4A\left(Lg,q_{\mathrm{max}}\right)q_{\mathrm{max}}^{3} \,.
\end{equation}
Accordingly, after equating, we obtain
\begin{equation}
A\left(Lg,q_{\mathrm{max}}\right)\approx Lg-\frac{3C}{4q_{\mathrm{max}}} \,,
\end{equation}
or in terms of discrete frequencies
\begin{equation}
A\left(g,n_{\mathrm{max}}\right)\approx g-\frac{3C}{4n_{\mathrm{max}}} \,. \label{eq:A heuristic}
\end{equation}
$A\left(g,n_{\mathrm{max}}\right)$ is thus found to grow linearly with $g$ and crucially, for small $g$, it is found to vanish and even becomes negative. The crossover frequency $n_c$ defined above in equation~(\ref{eq:crossover def}) occurs around
\begin{equation}
n_{c}\sim\left|\frac{C}{gn_{\mathrm{max}}-\frac{3}{4}C}\right|n_{\mathrm{max}} \,,
\end{equation}
thus if we take $C\sim 1$, we expect a heavily dominated quartic behaviour if $g\gtrsim 1$ and a heavily dominated cubic behaviour if $g\lesssim 2/n_{\mathrm{max}}$. For values between these extremes, neither the cubic contribution nor the quartic contribution can be neglected.

To summarise, we have found that, up to first order, the static structure factor in the nonlinear regime behaves like
\begin{equation}
\left\langle \bar{\xi}_{\vec{n}}\bar{\xi}_{-\vec{n}}\right\rangle ^{\left(1\right)}=\frac{1}{2\Gamma_{|\vec{n}|}}=\frac{1}{2n^{3}\left[C+A\left(g,n_{\mathrm{max}}\right)n\right]} \,,
\end{equation}
where $C\sim 1$ and $A\left(g,n_{\mathrm{max}}\right)$ is approximately given by equation~(\ref{eq:A heuristic}). In dimensional terms, this reads as
\begin{equation}
\left\langle \tilde{\xi}_{\vec{n}}\tilde{\xi}_{-\vec{n}}\right\rangle ^{\left(1\right)}=\frac{\sqrt{\left(2k_{\mathrm{B}}TL^{2}\right)/\left[\left(2\pi\right)^{4}Eh\right]}}{2n^{3}\left[C+A\left(g,n_{\mathrm{max}}\right)n\right]} \,.
\end{equation}

Note that the bending rigidity $\kappa$ does not enter the expression explicitly except via the dependence of $A\left(g,n_{\mathrm{max}}\right)$ on $g$. Applying the same logic as at the end of section~\ref{sec:linear static structure factor}, we find upon comparison with equation~(\ref{eq:zeta definition}) that the roughness exponent of the sheet in the nonlinear regime is $\zeta=1/2$ which is found to be in agreement with results presented in \cite{Nelson1987}. It is important however to emphasise that this roughness exponent only characterises the sheet well if we are in the extremely strong nonlinear regime of $g\lesssim2/n_{\mathrm{max}}$. When one finds oneself in the interval with $2/n_{\mathrm{max}}\lesssim g\lesssim1$, the existence of the ``linear'' quartic crossover cannot be neglected and thus the roughness exponent alone is less informative. Indeed, since $n_{\mathrm{max}}=L/\left(2a\right)$ where $a$ is the lattice length of the material, the extremely strong nonlinear regime of $g\lesssim2/n_{\mathrm{max}}$ will only be obtained if
\begin{equation}
\frac{\left(2\pi\right)^{2}\kappa}{\sqrt{2k_{\mathrm{B}}TEh\left(4a\right)^{2}}}\lesssim1 \,.
\end{equation}
For graphene, for instance, with a measured and theoretical bond-length of $a\sim0.15\textrm{ nm}$ \cite{Wei2013}, the left-hand side of this expression is actually roughly 8 and thus even graphene does not fall into this most extreme regime.

\subsection{The Dynamic Structure Factor\label{sec:nonlinear dynamic structure factor}}

As in the linear regime, we can obtain time-dependent two-point and four point functions by subbing $\mathbb{F}=\bar{\xi}_{\vec{n}_{1}}\left(0\right)\bar{\xi}_{\vec{n}_{2}}(\bar{t}\,)$ and $\mathbb{F}=\bar{\xi}_{\vec{n}_{1}}\left(0\right)\bar{\xi}_{\vec{\ell}_{1}}\bar{\xi}_{\vec{\ell}_{2}}\bar{\xi}_{\vec{\ell}_{3}}$ into equation~(\ref{eq:SCE}). At zeroth order ($m=0$), in complete analogy with the linear regime, this gives ODEs with solutions
\begin{equation}
\left\langle \bar{\xi}_{\vec{n}_{1}}\left(0\right)\bar{\xi}_{\vec{n}_{2}}(\bar{t}\,)\right\rangle ^{\left(0\right)}=\left\langle \bar{\xi}_{\vec{n}_{1}}\bar{\xi}_{\vec{n}_{2}}\right\rangle ^{\left(0\right)}e^{-\Gamma_{|\vec{n}_{2}|}\bar{t}}
\end{equation}
and
\begin{equation}
\left\langle \bar{\xi}_{\vec{n}_{1}}\left(0\right)\bar{\xi}_{\vec{\ell}_{1}}\bar{\xi}_{\vec{\ell}_{2}}\bar{\xi}_{\vec{\ell}_{3}}\right\rangle ^{\left(0\right)}=\left\langle \bar{\xi}_{\vec{n}_{1}}\bar{\xi}_{\vec{\ell}_{1}}\bar{\xi}_{\vec{\ell}_{2}}\bar{\xi}_{\vec{\ell}_{3}}\right\rangle ^{\left(0\right)}e^{-\Gamma_{|\vec{n}_{1}|}\bar{t}} \,,
\end{equation}
where the zeroth order equal-time two-point and four-point functions are given by equations~(\ref{eq:nonlin static 2 pt zeroth order}) and (\ref{eq:isserlis 2}).

We can obtain the first order correction to the dynamic two-point function by subbing $m=1$ and $\mathbb{F}=\bar{\xi}_{\vec{n}_{1}}\left(0\right)\bar{\xi}_{\vec{n}_{2}}(\bar{t}\,)$ into equation~(\ref{eq:SCE}). After simplification, we obtain the ODE
\begin{multline}
\frac{\partial\left\langle \bar{\xi}_{\vec{n}_{1}}\left(0\right)\bar{\xi}_{\vec{n}_{2}}(\bar{t}\,)\right\rangle ^{\left(1\right)}}{\partial\bar{t}}=-\Gamma_{|\vec{n}_{2}|}\left\langle \bar{\xi}_{\vec{n}_{1}}\left(0\right)\bar{\xi}_{\vec{n}_{2}}\right\rangle ^{\left(1\right)}\\
+\left[\Gamma_{|\vec{n}_{1}|} - g\left|\vec{n}_{1}\right|^{4} - \frac{1}{2}\sum_{\vec{\ell}\ne\vec{n}_{1}}\frac{\left|\vec{n}_{1}\times\vec{\ell}\,\right|^{4}}{\Gamma_{|\vec{\ell}\,|}\left|\vec{n}-\vec{\ell}\,\right|^{4}}\right]\left\langle \bar{\xi}_{\vec{n}_{1}}\bar{\xi}_{\vec{n}_{2}}\right\rangle ^{\left(0\right)}e^{-\Gamma_{|\vec{n}_{2}|}\bar{t}} \,.
\end{multline}

As pointed out previously, since the nonhomogeneous part of this equation decays at the natural decay rate of the equation, the solution to this equation will contain a nonphysical secular term unless the coefficient of the nonhomogeneous part happens to vanish. This will occur if
\begin{equation}
\Gamma_{|\vec{n}|}=g\left|\vec{n}\right|^{4}+\frac{1}{2}\sum_{\vec{\ell}\ne\vec{n}}\frac{\left|\vec{n}\times\vec{\ell}\,\right|^{4}}{\Gamma_{\left|\vec{\ell}\,\right|}\left|\vec{n}-\vec{\ell}\,\right|^{4}} \,,
\end{equation}
but this is none other than the discrete integral equation for $\Gamma_{|\vec{n}|}$ obtained in the previous subsection (equation~(\ref{eq:disc integ eq}))! Accordingly, all the results obtained for $\Gamma_{|\vec{n}|}$ in the previous subsection apply here as well and we thus find that the dynamic structure factor is given up to first order by an exponential decay with decay rate $\Gamma_{|\vec{n}|}$
\begin{equation}
\left\langle \bar{\xi}_{\vec{n}}\left(0\right)\bar{\xi}_{-\vec{n}}(\bar{t}\,)\right\rangle ^{\left(1\right)}=\left\langle \bar{\xi}_{\vec{n}}\bar{\xi}_{-\vec{n}}\right\rangle ^{\left(1\right)}e^{-\Gamma_{|\vec{n}|}\bar{t}} \,. \label{eq:nonlin dyn two pt function}
\end{equation}

The above argument that secular terms should be eliminated has already been alluded to in Section~\ref{sec:linear dynamic structure factor} when deriving the dynamic structure factor in the linear regime. There it was observed that the ordinary application of perturbation theory fails to eliminate the subsequent secular term and thus more sophisticated techniques such as the Poincar\'e-Lindstedt method or resummation via Pad\'e approximants must be used. Here in contrast, the SCE immediately provides a means of eliminating the secular term. Note that eliminating the secular term is conceptually equivalent to imposing that the zeroth order approximation be exact up to first order
\begin{equation}
\left\langle \bar{\xi}_{\vec{n}}\left(0\right)\bar{\xi}_{-\vec{n}}(\bar{t}\,)\right\rangle ^{\left(1\right)}=\left\langle \bar{\xi}_{\vec{n}}\left(0\right)\bar{\xi}_{-\vec{n}}(\bar{t}\,)\right\rangle ^{\left(0\right)}
\end{equation}
and thus can also be construed as a self-consistent argument for $\Gamma_{|\vec{n}|}$. 

The fact that the exact same integral equation is obtained for $\Gamma_{|\vec{n}|}$ from the analysis of the static and dynamic structure factors was by no means guaranteed and indeed, it is known to not be the case for the ordinary $\phi^4$-theory. Indeed, this fact suggests that the system is in some sense quasi-linear, i.e. it behaves like a purely linear theory with linear coupling given by $\Gamma_{|\vec{n}|}$. In dimensional terms, equation~(\ref{eq:nonlin dyn two pt function}) can be written as
\begin{equation}
\left\langle \tilde{\xi}_{\vec{n}}\left(0\right)\tilde{\xi}_{-\vec{n}}(t)\right\rangle ^{\left(1\right)}=\\ \left\langle \tilde{\xi}_{\vec{n}}\tilde{\xi}_{-\vec{n}}\right\rangle ^{\left(1\right)}e^{-\frac{\left(2\pi\right)^{2}\sqrt{2k_{\mathrm{B}}TEh}}{\alpha L^{3}}n^{3}\left[C+A\left(g,n_{\mathrm{max}}\right)n\right]t} \,,
\end{equation}
which makes it clear that at large scales, the dynamic exponent is simply $z=3$ which, together with our identification of $\zeta = 1/2$ for the nonlinear regime, indeed satisfies the relationship expected for a linear theory given by equation~(\ref{eq:exponent relationship}). It is important to note however that this quasi-linearity extends well beyond merely the large scale behaviour of the sheet characterised by these exponents. The fact that $\Gamma_{|\vec{n}|}$ has the same functional form for both the static and dynamic structures at all frequency scales in fact suggests that this quasi-linearity occurs at \textit{all} scales.

\section{Comparison with Simulations \label{sec:simulations}}

Equation~(\ref{eq:langevin equ dimen}) is a straightforward nonlinear Langevin equation and can thus be simulated using standard numerical techniques. First though, it will be helpful to nondimensionalise the equation. We do so using the nondimensionalisation used for the nonlinear analysis performed in Section~\ref{sec:nonlinear theory}, that is, using equations~(\ref{eq:t scaling}) and (\ref{eq:xi scaling}) while scaling the noise by
\begin{equation}
\bar{\eta}_{\vec{n}}\left(\bar{t}\right)=\left[\frac{L^{10}}{\left(2\pi\right)^{4}Eh\left(2k_{\mathrm{B}}T\right)^{3}}\right]^{1/4}\tilde{\eta}_{\vec{n}}\left(t\right) \,,
\end{equation}
we obtain the following dimensionless Langevin equation
\begin{equation}
\frac{\partial\bar{\xi}_{\vec{n}}}{\partial\bar{t}}=-g\left|\vec{n}\right|^{4}\bar{\xi}_{\vec{n}}\\
-\frac{1}{2}\sum_{\vec{\ell}_{1}\ne\vec{n}}\sum_{\vec{\ell}_{2}}\sum_{\vec{\ell}_{3}}V_{\vec{n},\vec{\ell}_{1},\vec{\ell}_{2},\vec{\ell}_{3}}\bar{\xi}_{\vec{\ell}_{1}}\bar{\xi}_{\vec{\ell}_{2}}\bar{\xi}_{\vec{\ell}_{3}}+\bar{\eta}_{\vec{n}}\left(\bar{t}\right) \,, \label{eq:langevin equ nondim}
\end{equation}
where the variance of the dimensionless noise is simply
\begin{equation}
\left\langle \bar{\eta}_{\vec{n}}\left(\bar{t}\right)\bar{\eta}_{\vec{n}'}\left(\bar{t}'\right)\right\rangle =\delta\left(\bar{t}-\bar{t}'\right)\delta_{\vec{n},-\vec{n}'} \,.
\end{equation}
Following \cite{Steinbock2022} and \cite{Steinbock2023}, for various values of $g$, we simulate equation~(\ref{eq:langevin equ nondim}) by forward iteration over a $41\times41$ square lattice implying a maximum frequency $n_\mathrm{max} = 20$. We do this for both large and small values of $g$ corresponding to the linear and nonlinear regimes respectively. To determine the dimensionless time step $\delta \bar{t}$ and time duration $\bar{T}$ of each simulation, we recall that equations~(\ref{eq:Gamma n final}) and (\ref{eq:A heuristic}) suggest that modes of size $n$ decay at rate
\begin{equation}
\Gamma_{n}\sim Cn^{3}+\left(g-\frac{3C}{4n_{\mathrm{max}}}\right)n^{4} \,,
\end{equation}
where $C\sim 1$ is relatively constant across values of $g$. Subbing in the minimum frequency $n = 1$ and maximum frequency $n = n_\mathrm{max}$, we obtain two time scales
\begin{equation}
\tau_{\mathrm{min}}\sim\frac{1}{\Gamma_{n_{\mathrm{max}}}}\sim\frac{4}{n_{\mathrm{max}}^{3}\left(1+4gn_{\mathrm{max}}\right)} \label{eq: tau min}
\end{equation}
and
\begin{equation}
\tau_{\mathrm{max}}\sim\frac{1}{\Gamma_{1}}\sim\frac{1}{1+g} \,.
\end{equation}

The fastest decaying modes decay with characteristic time $\tau_\mathrm{min}$ thus our time step $\delta \bar{t}$ must be smaller than this if we wish to faithfully simulate the equation. Accordingly, for each simulation, we select the time step $\delta \bar{t}$ to be equal to the largest power of $2$ smaller than $\tau_\mathrm{min}$ (i.e. $\delta \bar{t} = 2^{\max(k)} < \tau_\mathrm{min}$). It is worth noting that in practice, if the time step $\delta \bar{t}$ is larger than $\tau_\mathrm{min}$, the simulation becomes unstable and thus breaks down after a small number of steps. Accordingly, on a practical level, we can use the stability of the simulation to gauge whether our time step is sufficiently small.

The slowest decaying modes decay with characteristic time $\tau_\mathrm{max}$ thus if we wish to observe their decay, we must run the simulations over time intervals $\bar{T}$ larger than $\tau_\mathrm{max}$. Accordingly, we select the time duration of each simulation to be equal to $4$ times the smallest power of $2$ larger than $\tau_\mathrm{max}$ (i.e. $\bar{T}/4 = 2^{\min(k)} > \tau_\mathrm{max}$), the factor of $4$ being chosen to ensure that we have at least several independent intervals in which the decay is guaranteed to occur.

The aforementioned decision to use a $41\times41$ lattice, i.e. $n_\mathrm{max} = 20$, accounts for the competing interests of spatial resolution and temporal resolution in our simulation. Equation~(\ref{eq: tau min}) makes it clear that finer spatial resolution (i.e. larger values of $n_\mathrm{max}$) requires substantially finer temporal resolution. Since we wish to be able to run the simulations for long enough to obtain statistics for even the slowest decaying modes, using a temporal resolution which is extremely fine is computationally expensive. Running simulations with different values of $n_\mathrm{max}$, it was observed that $n_\mathrm{max} = 20$ handles this trade-off satisfactorily, that is, the spatial resolution is sufficiently fine for our theory to be applicable while the temporal resolution is not too fine to make the simulations too onerous to run.

Finally, since it is time consuming to calculate the nonlinear quartic term of equation~(\ref{eq:langevin equ nondim}) directly, we implement the pseudo-spectral method described in \cite{During2017} and used in \cite{Steinbock2022} and \cite{Steinbock2023} in which the term is calculated as the Fourier transform of its real-space counterpart. This has the additional effect of imposing periodic boundary conditions on the simulation. For each value of $g$ used, 10 simulations were run and the results averaged across simulations.

\begin{figure}
\centering
\includegraphics[width=11cm]{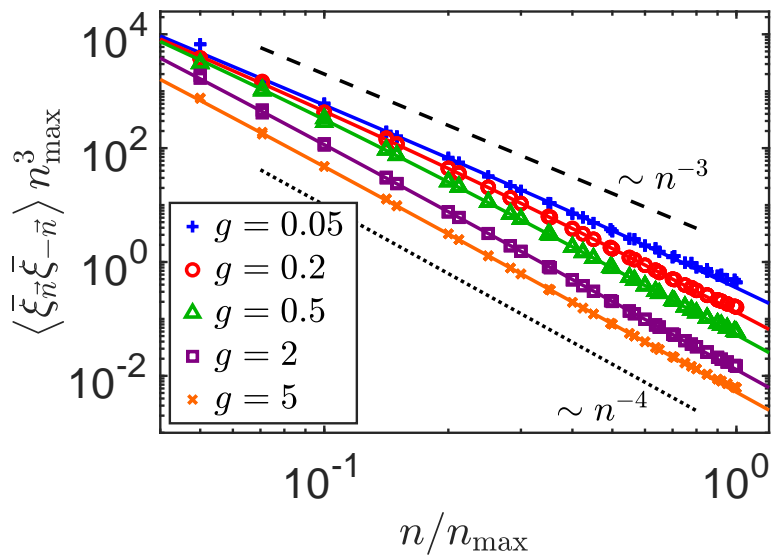}%
\caption{The static structure factor $\left\langle \bar{\xi}_{\vec{n}}\bar{\xi}_{-\vec{n}}\right\rangle$ as a function of $n$ for various values of $g$. The data points (symbols) denote simulation data while the solid lines denote fits of equation~(\ref{eq:static fit}) to this data. The dashed and dotted lines are guidelines proportional to $n^{-3}$ and $n^{-4}$ respectively. For large values of $g$, the $n^{-4}$ behaviour dominates, confirming the prediction of the linear theory while for small values of $g$, the $n^{-3}$ behaviour dominates, confirming the prediction of the nonlinear theory. For intermediate values, a crossover between $n^{-3}$ behaviour for small $n$ and $n^{-4}$ behaviour for large $n$ is observed.
\label{fig:static 2 pt}}
\end{figure}

Figure~\ref{fig:static 2 pt} shows the static structure factor for several values of $g$ ranging from significantly smaller than unity to significantly larger than unity. The points (symbols) denote the simulation data while the solid lines denote fits to the data of the equation
\begin{equation}
\left\langle \bar{\xi}_{\vec{n}}\bar{\xi}_{-\vec{n}}\right\rangle =\frac{1}{2\left(Cn^{3}+An^{4}\right)} \,, \label{eq:static fit}
\end{equation}
where $A$ and $C$ are treated as fitting parameters. The dashed and dotted lines are guidelines proportional to $n^{-3}$ and $n^{-4}$ respectively. As can be seen, as $g$ increases, the simulation data moves from being roughly parallel to the $n^{-3}$ guideline, predicted for the theory of the nonlinear regime, to being roughly parallel to the $n^{-4}$ guideline, predicted for the theory of the linear regime. For intermediate values of $g$, a clear crossover between $n^{-3}$ behaviour for small $n$ and $n^{-4}$ behaviour for large $n$ is observed. The fact that the fits are excellent across all values of $n$ for all values of $g$ confirms that equation~(\ref{eq:static fit}) provides the correct functional form of the static structure factor. The fit parameters $C(g)$ and $A(g)$ are plotted in figure~\ref{fig:fit parameters} and will be discussed in the continuation.

Before moving to the dynamic structure factor, it is worth observing that it is easy to naively fit power-laws $n^{-\nu}$ to the data presented in figure~\ref{fig:static 2 pt} and this will give values of $\nu$ between 3 and 4. Indeed, because the change from $n^{-3}$ behaviour to $n^{-4}$ behaviour is visually quite subtle, this will even be the case when one only attempts to fit the small $n$ values. Previous research has indeed observed a variety of anomalous scaling exponents between 3 and 4 \cite{Kantor1986, Kantor1987, Kardar1987, Aronovitz1989, Bowick1996, LeDoussal1992, NelsonBook2004, Fasolino2007, Los2009, Troster2013a, Troster2013b, Mizuochi2014, Los2016, LeDoussal2018, Granato2022} and even observed a temperature dependence \cite{Thomas2015} despite claims of universality. A similar state of affairs was previously observed in the context of fracture where a range of exponents could be extracted from naive power-law fitting of data that was known to be exactly described by a rational function, that is, a ratio of two polynomials \cite{Katzav2007b, Katzav2013}. Here too, we suggest that the range of various anomalous exponents is merely an artefact of the effect of the large $n$ linear behaviour on the small $n$ leading nonlinear character.

\begin{figure}
\centering
\includegraphics[width=11cm]{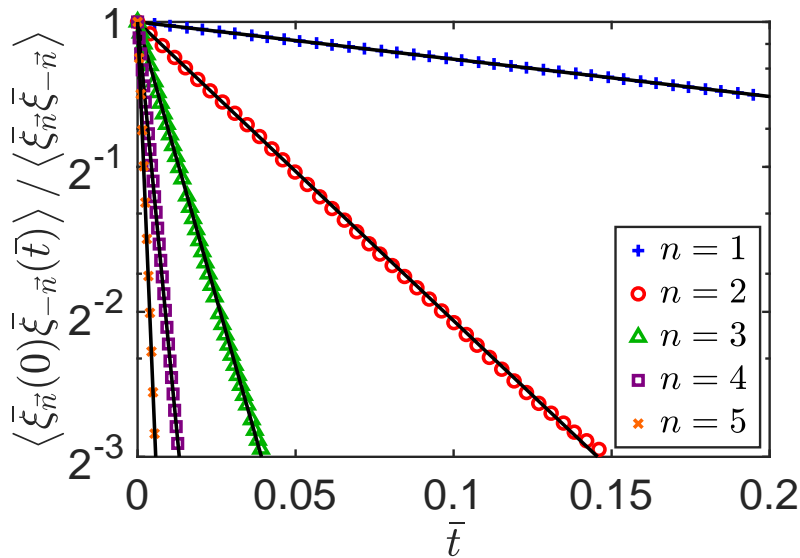}%
\caption{The time-dependent structure factor $\left\langle \bar{\xi}_{\vec{n}}(0)\bar{\xi}_{-\vec{n}}(\bar{t})\right\rangle$ as a function of $\bar{t}$ for $n=1,2,3,4,5$ when $g=0.5$. The solid lines denote fits of exponential decays to each data set. Since the data appears to fall on straight-lines on the logarithmic y-axis, we see that the time-dependent two-point function indeed decays exponentially, as predicted by the theory.
\label{fig:decaying 2 pt}}
\end{figure}

\begin{figure}
\centering
\includegraphics[width=11cm]{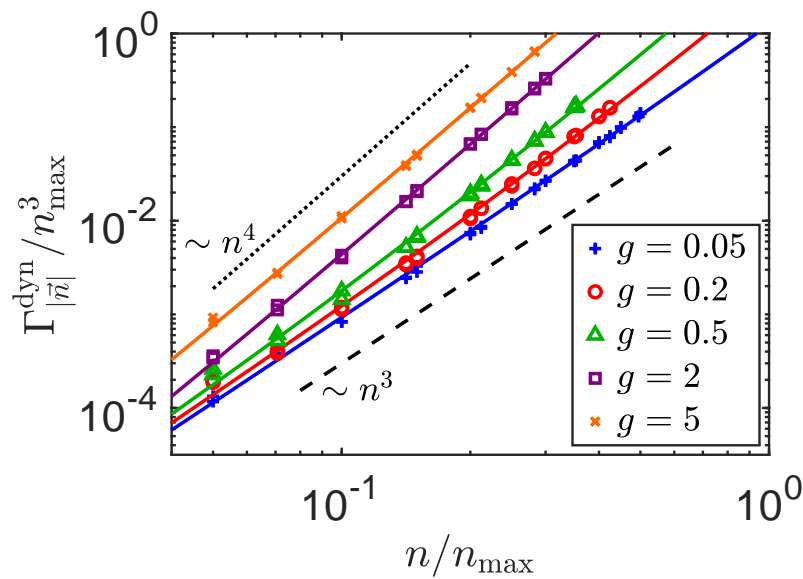}%
\caption{The exponential decay rate $\Gamma_{|\vec{n}|}^{\mathrm{dyn}}$ of the time-dependent structure factor as a function of $n$ for various values of $g$. The data points denote the simulation data while the solid lines denote fits of equation~(\ref{eq:Gamma fit}) to the data. The dashed and dotted lines are guidelines proportional to $n^3$ and $n^4$ respectively. For large values of $g$, the $n^{4}$ behaviour dominates, confirming the prediction of the linear theory while for small values of $g$, the $n^{3}$ behaviour dominates, confirming the prediction of the nonlinear theory. For intermediate values, a crossover between $n^{3}$ behaviour for small $n$ and $n^{4}$ behaviour for large $n$ is observed.
\label{fig:Gamma dyn}}
\end{figure}

To confirm our predictions for the dynamical properties of the system, the time-dependent structure factor can be plotted as a function of $\bar{t}$ and fitted by an exponential decay from which the decay rate $\Gamma_{|\vec{n}|}^{\mathrm{dyn}}$ can be extracted as suggested by equation~(\ref{eq:nonlin dyn two pt function}). Figure~\ref{fig:decaying 2 pt} shows this decay for several values of $n$ for $g=0.5$ (safely in the nonlinear regime) with the solid lines denoting fits of exponential decays to each data set. Since the y-scale is logarithmic, the decays appear as straight lines confirming our theory that indeed, the time-dependent structure factor decays exponentially with time. Similar results can be obtained for other values of $n$ and $g$ and the results are presented in figure~\ref{fig:Gamma dyn} (symbols) together with fits (solid lines) of the equation
\begin{equation}
\Gamma_{\left|\vec{n}\right|}^{\mathrm{dyn}}=Cn^{3}+An^{4} \,. \label{eq:Gamma fit}
\end{equation}
The dashed and dotted lines are guidelines proportional to $n^3$ and $n^4$ respectively and it can be seen that as $g$ increases, the behaviour of the data shifts from being parallel to the $n^3$ guideline, predicted for the nonlinear theory, to being parallel to the $n^4$ guideline, predicted by the linear theory. Indeed, the fact that the fits of equation~(\ref{eq:Gamma fit}) are excellent over all values of $n$ for all values of $g$ confirms that equation~(\ref{eq:Gamma fit}) correctly predicts the functional form of the decay rate of the time-dependent structure factor. The fit parameters $C(g)$ and $A(g)$ are plotted in figure~\ref{fig:fit parameters} and will be discussed shortly. 

\begin{figure}
\centering
\includegraphics[width=11cm]{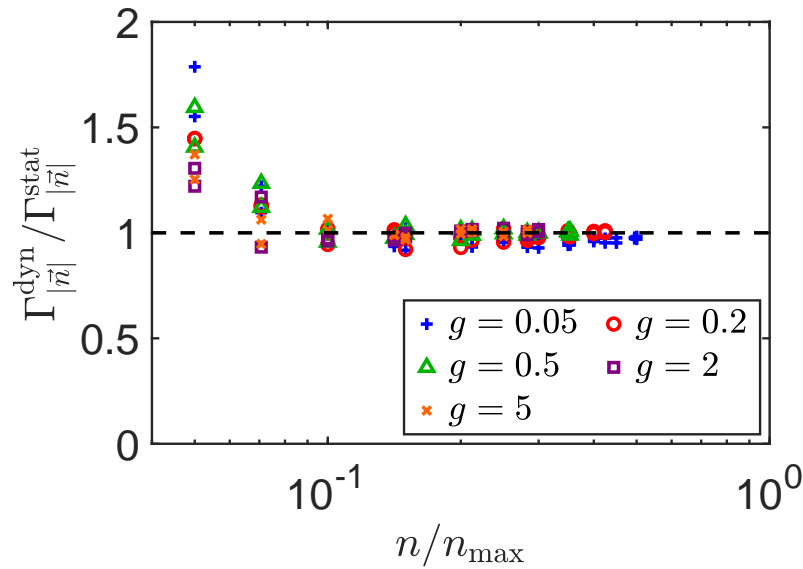}%
\caption{The ratio of the decay rate of the time-dependent structure factor, $\Gamma_{|\vec{n}|}^{\mathrm{dyn}}$, to the effective linear coupling, $\Gamma_{|\vec{n}|}^{\mathrm{stat}}$, given by equation~(\ref{eq:Gamma static}) as a function $n$ for various $g$. The ratio can be seen to be roughly equal to 1 for all values of $g$ except when $n$ becomes small at which point the ratio increases slightly.
\label{fig:Gamma ratio}}
\end{figure}

In addition to predicting the functional forms of the static structure factor (equation~(\ref{eq:static fit})) and the decay rate of the time-dependent structure factor (equation~(\ref{eq:Gamma fit})), our theory predicts that the system is quasilinear at all scales, that is to say, that the effective linear coupling $\Gamma_{\left|\vec{n}\right|}^{\mathrm{stat}}$, which can be defined using
\begin{equation}
\left\langle \bar{\xi}_{\vec{n}}\bar{\xi}_{-\vec{n}}\right\rangle =\frac{1}{2\Gamma_{\left|\vec{n}\right|}^{\mathrm{stat}}} \,, \label{eq:Gamma static}
\end{equation}
should equal the decay rate of the time-dependent structure factor $\Gamma_{\left|\vec{n}\right|}^{\mathrm{dyn}}$, i.e. we should have
\begin{equation}
\frac{\Gamma_{\left|\vec{n}\right|}^{\mathrm{dyn}}}{\Gamma_{\left|\vec{n}\right|}^{\mathrm{stat}}}=1
\end{equation}
for all $\vec{n}$ and $g$. Figure~\ref{fig:Gamma ratio} shows this ratio for various values of $g$ as a function of $n$ and indeed, for all $g$, this ratio is close to 1 except when $\vec{n}$ is small at which point the ratio becomes slightly yet noticeably larger than 1. The fact this ratio is close to 1 for most $n$ shows that our prediction that the system is quasi-linear is indeed mostly correct. For large $g$, we expect the system to become increasingly linear and indeed, the deviation for small $n$ appears to shrink with increasing $g$. The deviation itself is surprising considering the fact that figure~\ref{fig:static 2 pt} and figure~\ref{fig:Gamma dyn} each individually show excellent agreement across all values of $n$. We are thus forced to conclude that though equations~(\ref{eq:static fit}) and (\ref{eq:Gamma fit}) are each functionally correct, the fit parameters $C(g)$ and $A(g)$ are not necessarily identical. 

\begin{figure}
\centering
\includegraphics[width=8cm]{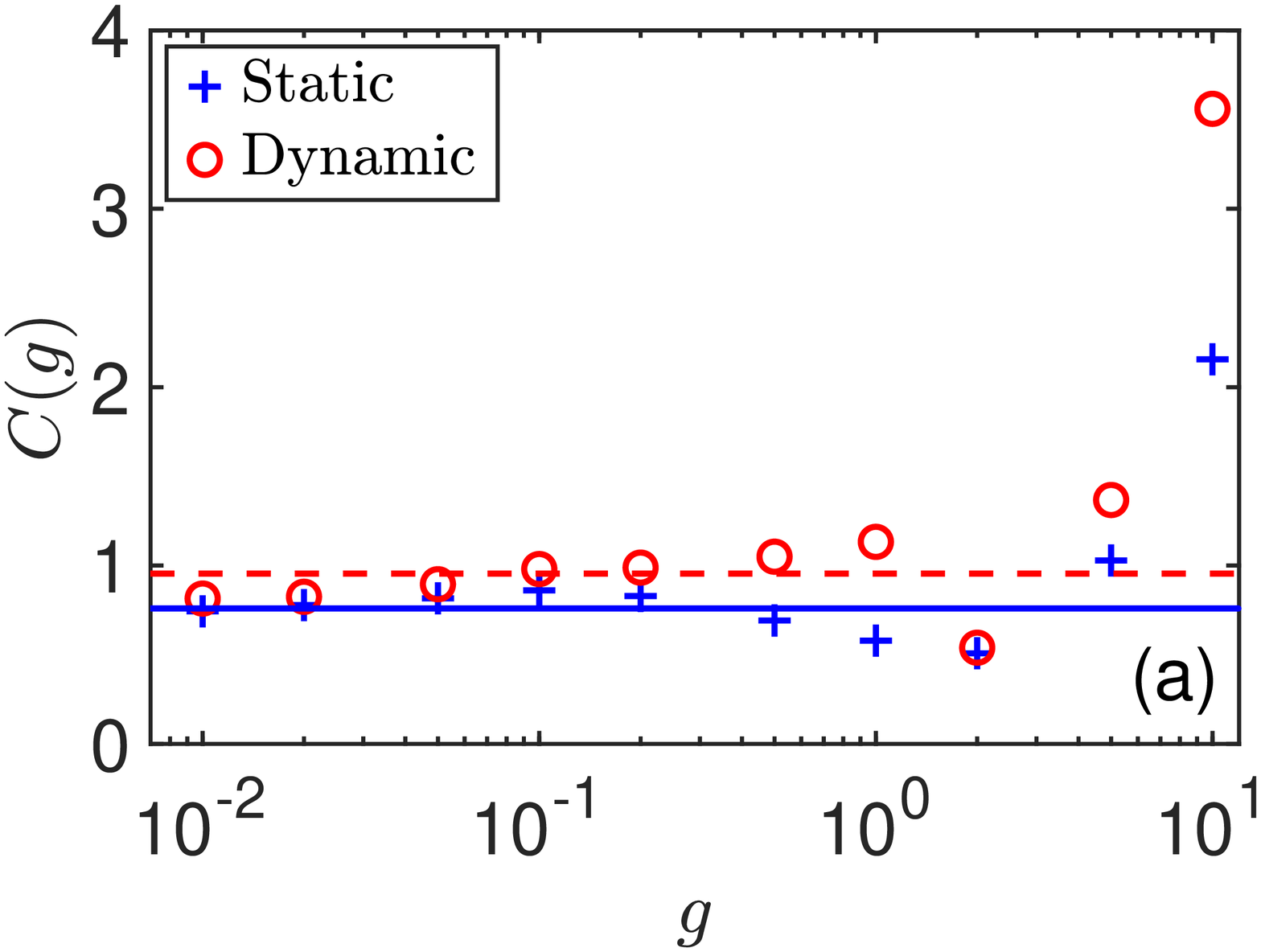}%
\includegraphics[width=8cm]{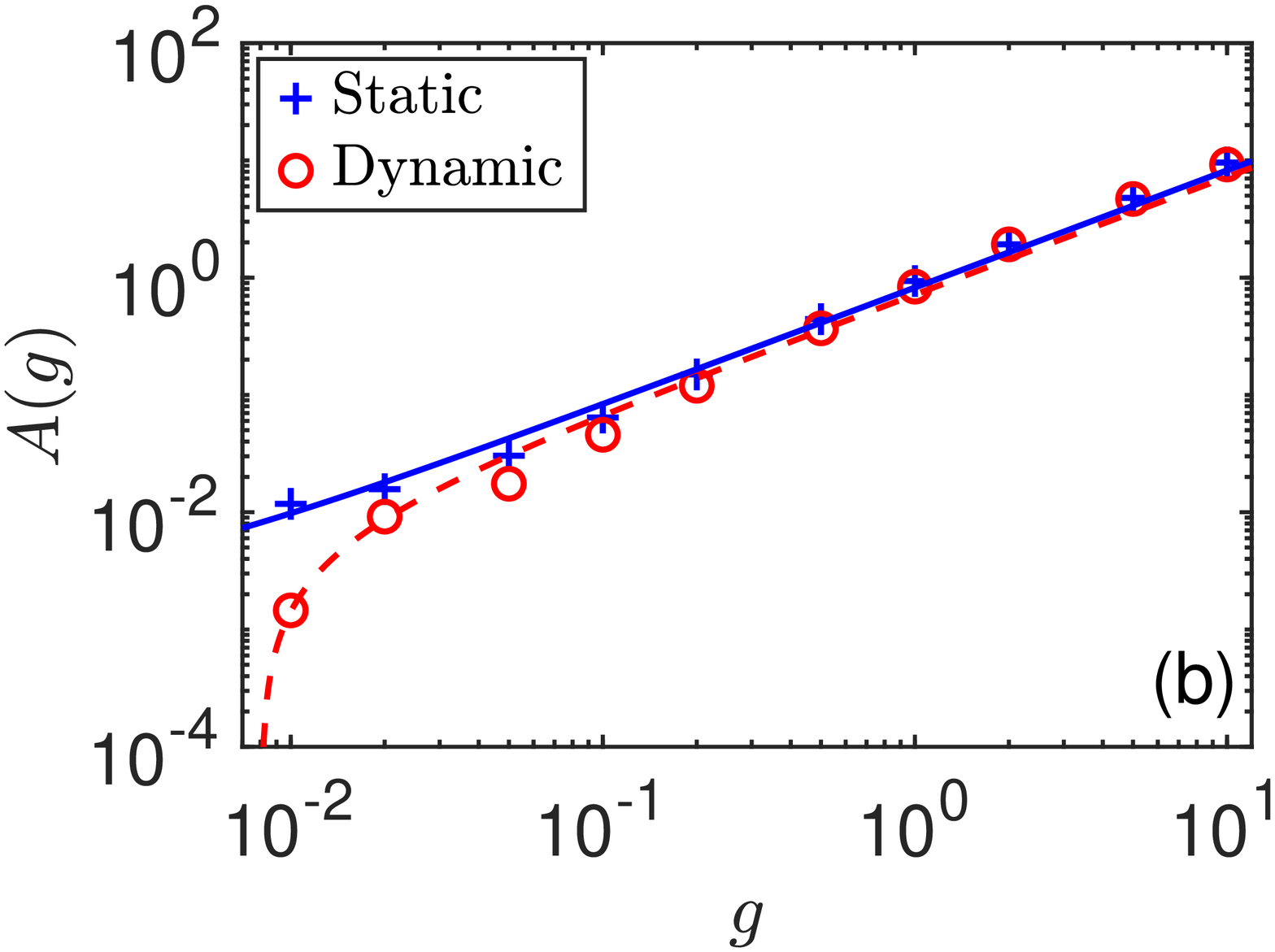}
\caption{The fit parameters, (a) $C(g)$, (b) $A(g)$, obtained by fitting equations~(\ref{eq:static fit}) and (\ref{eq:Gamma fit}) to the static data presented in figure~\ref{fig:static 2 pt} (blue +) and the dynamic data presented in figure~\ref{fig:Gamma dyn} (red $\circ$) respectively. In (a), the solid and dashed lines are the average value of the static and dynamic $C(g)$ respectively. In (b), the solid and dashed lines are linear fits to the static and dynamic $A(g)$ respectively. 
\label{fig:fit parameters}}
\end{figure}

Figure~\ref{fig:fit parameters} shows the fit parameters $C(g)$ and $A(g)$ as functions of $g$ with the static values (blue +) coming from fits of equation~(\ref{eq:static fit}) to the data displayed in figure~\ref{fig:static 2 pt} and the dynamic values (red $\circ$) coming from fits of equation~(\ref{eq:Gamma fit}) to the data displayed in figure~\ref{fig:Gamma dyn}. As predicted by the theory, the values of $C(g)$ remain relatively constant while the values of $A(g)$ grow with increasing $g$. 

Since the theory predicts that $C(g)$ is relatively constant, the solid and dashed lines in figure~\ref{fig:fit parameters}(a) denote the average value of $C(g)$ obtained from the static and dynamic data respectively and, as can be seen, the deviations around this average value are indeed small over the entire range of $g$ except when $g$ becomes large. Our theory predicts that $C(g)\sim1$ and this too is indeed the case for most values of $g$ for both sets of $C(g)$. Since our theory predicts that for large $g$, $\Gamma_{|\vec{n}|}$ will grow mostly like $n^4$, the sudden growth in $C(g)$ when $g$ is large is merely an artefact of the result that $C(g)$ isn't needed as a fit parameter in this regime.

According to equation~(\ref{eq:A heuristic}), $A(g)$ is expected to grow linearly. The solid and dashed lines in figure~\ref{fig:fit parameters}(b) denote linear fits to the $A(g)$ obtained from the static and dynamic data respectively and can be seen to be excellent over all values of $g$. It is worth noting that according to equation~(\ref{eq:A heuristic}), $A(g)$ is expected to become negative for sufficiently small values of $g$ and this behaviour is clearly visible on the log-log plot for the dynamic $A(g)$ where the dashed line suddenly plummets to $-\infty$. Indeed, the static $A(g)$ and dynamic $A(g)$ are almost identical for large values of $g$ and only really begin to differ for small values of $g$ with the dynamic $A(g)$ becoming negative while the static $A(g)$ remains positive. 

Accordingly, it seems that the small $n$ deviation visible in figure~\ref{fig:Gamma ratio} is primarily due to the small discrepancy between the values of $A(g)$ obtained from the static and dynamic data. This discrepancy in the coefficient of the quartic behaviour, indicating that the quasi-linearity is not perfect, is completely unpredicted by our theory though it is worth emphasising that it only appears to become significant for small values of $g$. Conceptually, we seem to have found that the quartic behaviour of the decay rate of the time-dependent two-point function is a renormalisation of the quartic behaviour of the effective linear coupling. Properly accounting for this renormalisation is left as an open question.

\section{Discussion\label{sec:discussion}}

We have used the formalism of statistical mechanics to develop a Langevin equation describing an elastic thin sheet being driven by white or thermal noise. By investigating this so-called dynamic F\"oppl-von K\'arm\'an equation, we have found that the behaviour of such a sheet can be characterised as belonging to one of three ``classes'' depending on the value of the single dimensionless parameter $g$, given by equation~(\ref{eq:g definition}). For $g\gg1$, the system resides in the mostly linear regime where the sheet's curvature plays little role and thus the equal-time structure factor in Fourier space $\langle \tilde{\xi}_{\vec{n}}\tilde{\xi}_{-\vec{n}}\rangle $ is found to decay like $\sim n^{-4}$ with corresponding roughness exponent $\zeta = 1$. For $g\ll 2/n_{\textrm{max}}$, the system resides in the highly nonlinear regime where the effects of curvature dominate at all scales and thus, consistent with previously analyses \cite{Nelson1987, Ahmadpoor2017}, the equal-time structure factor is renormalised to decay like $\sim n^{-3}$ with corresponding roughness exponent $\zeta = 1/2$. In between these two extremes, an intermediate behaviour is obtained where the equal-time structure factor is found to decay like $\sim n^{-3}$ for small frequencies (large scales) and $\sim n^{-4}$ for large frequencies (small scales). This intermediate scale is of particular interest as it seems likely that many ultra-thin, novel two-dimensional materials such as graphene lie within it.

Indeed, there is much research indicating that for thin and ultra-thin materials, the equal-time structure factor decays with an exponent between 3 and 4 \cite{Kantor1986, Kantor1987, Kardar1987, Aronovitz1989, Bowick1996, LeDoussal1992, NelsonBook2004, Fasolino2007, Los2009, Troster2013a, Troster2013b, Mizuochi2014, Los2016, LeDoussal2018, Granato2022}, with corresponding roughness exponent $\zeta$ between 1/2 and 1, or is even temperature dependent \cite{Thomas2015}. Here, we argue that many of these results can be better explained as insufficiently accounting for the crossover to the linear tail. The transition from an $n^{-3}$ behaviour to an $n^{-4}$ behaviour can be quite subtle and misidentifying the location of the crossover by even a small amount can cause naive power-law fits to produce anomalous exponents. A similar phenomenon is reported in the context of fracture in references \cite{Katzav2007b, Katzav2013}.  The confirmation of our results provided by our simulations supports this position.

In addition to recovering and extending the predicted theory for the static structure factor, we have used our model to provide the first concrete prediction for the dynamic structure factor $\langle \tilde{\xi}_{\vec{n}}\left(0\right)\tilde{\xi}_{-\vec{n}}\left(t\right)\rangle $, finding analytic predictions in all regimes for the exponential decay towards equilibrium and predicting a dynamic exponent $z$ given by $z = 2\zeta + 2$. Importantly, this relationship between the dynamic exponent $z$ and the roughness exponent $\zeta$ holds in all regimes and thus even in the most extreme nonlinear regime, the model we have developed can be thought of as quasi-linear. 

This quasi-linearity can to some extent be interpreted as a consequence of a pair of known exponent inequalities \cite{Katzav2011a, Katzav2011b}. In particular, it is known that for systems whose dynamics can be derived from a Hamiltonian, that the dynamic exponent $z$ must obey the inequality $z \ge 2\zeta + d$ where $d$ denotes the dimension of the system and is thus equal to 2 in our case. Simultaneously, it is known that Galilean invariant systems, i.e. systems for which the dynamics of higher modes do not depend on the behaviour of the zeroth mode $\tilde{\xi}_0$, must obey the reverse inequality $z \le 2\zeta + d$. Since our system can indeed be derived from a Hamiltonian \cite{LandauLifshitzElasticityBook, NelsonBook2004, During2017, Ahmadpoor2017} and has been observed to be manifestly Galilean invariant [see the discussion immediately following equation~(\ref{eq: zeroth mode equ})], these two inequalities collapse into the single equality $z = 2\zeta + d$, independent of the size of the nonlinear coupling. From this perspective, it is an important and nontrivial example of a physical system where both inequalities apply. It is worth noting however that the application of these inequalities is limited in scope to only the largest scales while we have found in our particular system that the quasi-linearity applies at \textit{all} scales. That is, the exponent inequalities mentioned in \cite{Katzav2011a, Katzav2011b} only guarantee that the leading order powers $\zeta$  and $z$ will be related by $z = 2\zeta + d$ while we have found that the prefactors and subleading orders also coincide. Indeed, our simulations show that the \textit{only} significant deviation from quasi-linear behaviour is in the large frequency, small scale coefficient $A(g)$ and thus the exponent inequalities are not contradicted. Accordingly, this quasi-linearity across all scales regardless of the size of the nonlinear coupling is both unexpected and highly significant. To the best of our knowledge, a theory which could \textit{a priori} predict this broad degree of quasi-linearity directly from the structure of the equations is lacking despite its obvious great value.

It is also worth mentioning that in fact, not only do the higher modes not depend on the zeroth mode (Galilean invariance) but the zeroth mode also does not depend on the higher modes. Accordingly, the sheet's center of mass performs a random walk in space without being affected by the structure of the sheet, while also not affecting its structure. This prediction is amenable to experimental test and could help decipher the dynamics of the sheet. In particular, if no such decoupling is observed, or if the stochastic motion of the center of mass exhibits a different pattern, this could be valuable information. 

The results discussed here could also have an interesting application to the optical response of thin sheets, or actually any waves interacting with the structure, including acoustic waves. In particular, crumpled papers are known to exhibit speckle patterns \cite{Rad2019} and to emit a well-distinguishable crackling noise \cite{Kramer1996, Houle1996}. Here too, it would be interesting to see if optical effects could be observed in graphene, which might be of practical use.

Finally, while the focus of this work has been on understanding the structure of thin sheets by studying the impact of thermal fluctuations on their curvature, there are many other curvature mediated means of effecting the structure of thin matter. For instance, it is known that the surface composition of thin films also modifies their bending moduli in a manner not dissimilar to that of thermal fluctuations \cite{Safran1999}. The equilibrium composition of polymer-blend films, however, remains only partially understood (see for instance \cite{Pellicane2016}). Accordingly, it would be interesting to ask whether the theory and methods developed here could provide insight into other types of curvature-structure interactions.

\ack
The authors wish to thank Arezki Boudaoud for useful discussions.
This work was supported by the Israel Science Foundation Grant No. 1682/18.

\printbibliography

\end{document}